\documentclass[twocolumn]{aa}  
\usepackage{graphicx}
\usepackage{txfonts}
\newcommand{\Msun}{\hbox{$\hbox{M}_\odot\;$}}

\newcommand{\kms}{\hbox{${\rm km}\:{\rm s}^{-1}\;$}}
\newcommand{\Msuno}{\hbox{$\hbox{M}_\odot$}}

\newcommand{\kmso}{\hbox{${\rm km}\:{\rm s}^{-1}$}}
\newcommand{\teff}{$T_{\rm eff}\;$}  
\newcommand{\teffo}{$T_{\rm eff}$}  
\newcommand{\logg}{$\log\;g\;$}  
\newcommand{\loggo}{$\log\;g$}

\begin{document}
   \title{The Black Hole Binary Nova Scorpii 1994 (\mbox{GRO
   J1655-40}):}

   \subtitle{An improved chemical analysis\thanks{Based
   on observations obtained with UVES at VLT Kueyen 8.2 m telescope in
   programme 073.D-0473(A)}}

   \author{J. I. Gonz\'alez Hern\'andez\inst{1,2,3},  R.
   Rebolo\inst{3,4},\and G. Israelian\inst{3}}

   \offprints{J. I. Gonz\'alez Hern\'andez}

   \institute{Observatoire de Paris-Meudon, GEPI, 5 place Jules
   Janssen, 92195 Meudon Cedex, France \\
   \email{}Jonay.Gonzalez-Hernandez@obspm.fr 
         \and
   CIFIST Marie Curie Excellence Team
         \and
   Instituto de Astrof{\'\i }sica de Canarias, E-38205 La Laguna,
   Tenerife, Spain \\ \email{}rrl@iac.es, gil@iac.es 
         \and
   Consejo Superior de Investigaciones Cient{\'\i }ficas, Spain
}

   \date{Received January, 2007; accepted --}

 
  \abstract
   {The chemical analysis of secondary stars of low mass X-ray binaries
   provides an opportunity to study the formation processes of
   compact objects, either black holes or neutron stars.}   
   {Following the discovery of overabundances of $\alpha$-elements in
   the Keck~I/HIRES spectrum of the secondary star of Nova Scorpii 1994
   (Israelian et al.\ 1999), we obtained VLT/UVES high-resolution
   spectroscopy with the aim of performing a detailed abundance
   analysis of this secondary star.}
   {Using a $\chi2$-minimization procedure and a grid of
   synthetic spectra, we derive the stellar parameters and atmospheric 
   abundances of O, Mg, Al, Ca, Ti, Fe and Ni, using a new UVES
   spectrum and the HIRES spectrum.}
   {The abundances of Al, Ca, Ti, Fe and Ni seem to be consistent with
   solar values, whereas Na, and especially O, Mg, Si and S are
   significantly enhanced in comparison with Galactic trends of these
   elements. A comparison
   with spherically and non-spherically symmetric supernova explosion
   models may provide stringent constraints to the model parameters as
   mass-cut and the explosion energy, in particular from the relative
   abundances of Si, S, Ca, Ti, Fe and Ni.} 
   {Most probably the black hole in this system formed in a hypernova
   explosion of a 30--35 \Msun progenitor star with a mass-cut in the
   range 2--3.5 \Msun. However, these models produce abundances of Al
   and Na almost ten times higher than the observed values.} 


   \maketitle
%

\section{Introduction}

The low mass X-ray binary Nova Scorpii 1994 (\mbox{GRO
J1655--40}) is one of the black hole transients that has attracted more
attention in the last decade. It was discovered on 1994 July 27 with
BATSE on board the \emph{Compton Gamma Ray Observatory} (Zhang et al.\
1994). The strong evidence that the compact object is a black hole was
presented by Bailyn et al. (1995) who suggested a mass function of
$f(M) = 3.16 \pm 0.15$ \Msuno. A self-consistent analysis of the
ellipsoidal light curves of the system has yielded masses of
$5.4\pm0.3$ and $1.45\pm0.35$ \Msun for the black hole and
the companion star (Beer \& Podsiadlowski 2002). 

Israelian et al. (1999) found evidence of the formation of
the black hole in an explosive supernova (SN) event. Using a high-resolution
spectrum obtained with the Keck~I Telescope, they found that the
atmosphere of the secondary was enriched by a factor of 6--10 in
several $\alpha$-process elements (O, Mg, Si and S). 
These elements, almost exclusively synthesized during supernova
explosions, cannot be produced in a low-mass secondary star which must
have been exposed to the supernova 
material ejected when the compact remnant was formed. Subsequent
discussions on the large amount of S, O, Mg and Si have considered
this object as a relic of a hypernova and Gamma Ray Burst (Brown et
al.\ 2000). More recently, these abundance anomalies were compared
with a variety of SN models, including standard as well as hypernova
models (for several progenitor masses and geometries), and a simple
model of the evolution of the binary and the pollution of the
secondary (Podsiadlowski et al.\ 2002). The observed anomalies were
explained for a large range of model parameters, providing substantial
fallback and mixing between the fallback matter and the ejecta. 

Mirabel et al. (2002) found additional independent evidence that
the black hole in this system formed in a SN explosion. They
determined the proper motion of the system using the \emph{Hubble
Space Telescope} (HST) and computed its galactocentric orbit. The
system moves with a runaway velocity of $\sim 112 \pm 18$ \kms in a
highly eccentric galactic orbit.

Here we improve the chemical analysis of the secondary star including
more elements like Na, Al, Ca, Ni and Li and re\-fi\-ning the abundance
measurements of elements already studied by Israelian et al. (1999).

%

\section{Observations}

We obtained 14 spectra of \mbox{Nova Sco 94} with the UV-Visual
Echelle Spectrograph (UVES) at the European Southern Observatory
(ESO), {\itshape Observatorio Cerro Paranal}, using the 8.2 m {Very
Large Telescope} (VLT) on 2000 April 16 and 2000 June 18, covering the
spectral regions $\lambda\lambda$3300--4500 {\AA},
$\lambda\lambda$4800--5800 {\AA} and $\lambda\lambda$5800--6800 {\AA} 
at resolving power $\lambda/\delta\lambda\sim43,000$. The total
exposure time was 5.6 hours. The spectra were reduced in a standard
manner using the UVES reduction package within the MIDAS environment.
In this paper, we will also make use of one spectrum taken with
the HIRES spectrograph at Keck telescope, covering the spectral region
$\lambda\lambda$6800--8800 {\AA}, previously used by Israelian et al.
(1999). This near-IR spectrum has been boxcar-smoothed in wavelength 
with steps of 0.3 {\AA}, providing a S/N$\sim$45--50 in the
continuum.

\begin{figure}[ht!]
\centering
\includegraphics[width=6.5cm,angle=90]{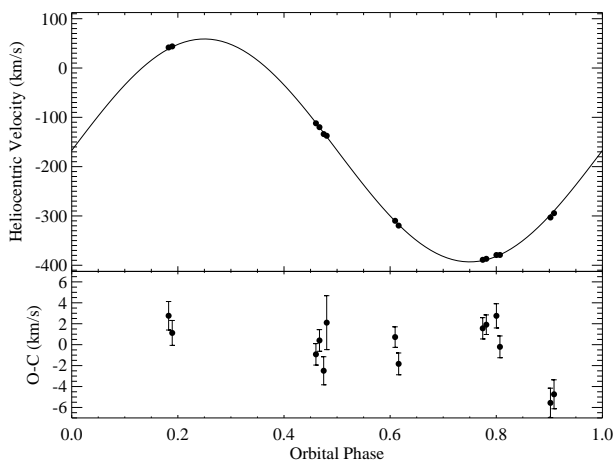}
\caption{\footnotesize{Radial velocities of \mbox{Nova Scorpii 1994}
folded on the orbital solution of the data with best fitting
sinusoid. Individual velocity errors are $\lesssim 1.5$ ${\rm
km}\:{\rm s}^{-1}$ and are not plotted because they are always 
smaller than the symbol size. The bottom panel shows the residuals
of the fit.}}   
\label{fig1}
\end{figure}

\subsection{Revised orbital parameters}

We extracted the radial velocities by cross-correlating each UVES
spectrum of the target with the spectrum of a F6IV template 
star (Shahbaz et al. 1999) taken from S$^4$N database (HIP 46853,
Allende Prieto et al. 2004), using the software {\scshape MOLLY}
developed by T. R. Marsh. A $\chi^2$ sine wave fit to the obtained
velocities yields to the following orbital solution (see
Fig.~\ref{fig1}): 
$\gamma=-167.1\pm0.6$ \kmso, $K_2=226.1\pm0.8$ \kmso,
$P=2.62120\pm0.00014$ d, and $T_0=2453110.5637\pm0.0019$ d, where
$T_0$ is defined as the corresponding time of the closest inferior
conjunction of the companion star, and the quoted uncertainties are at
1$\sigma$. This orbital period, $P$, together with the velocity
amplitude of the orbital motion of the secondary star, $K_2$, leads to
a mass function of $f(M)=3.16\pm0.03$ \Msun, consistent with previous
results (Bailyn et al. 1995; Orosz \& Bailyn 1997).

The derived radial velocity of the center of mass of the
system disagrees somewhat (at the 3$\sigma$ level) with previous
studies ($\gamma=-150\pm19$ \kmso, Bailyn et al. 1995;
$\gamma=-142.4\pm1.6$ \kmso, Orosz \& Bailyn 1997; 
$\gamma=-141.9\pm1.3$ \kmso, Shahbaz et al. 1999). Note that our
high-resolution data has a factor greater than 5 better spectral
resolution than the data these authors used. However, Shahbaz et al.
(1999) estimate the uncertainty in their individual radial velocity
measurements at typically 6 \kms, whereas our individual measurements
have error bars of roughly 1\kmso. In addition, we have checked the
accuracy of the wavelength calibration using the $\lambda$6300.2
[\ion{O}{i}] sky line but unfortunately, its line profile appears
distorted in each individual spectra of the target probably due to
stellar lines, providing an estimate of its line center within 3
\kmso. These uncertainties seem not to be enough to explain the difference between
both radial velocity measurements. 

\begin{figure*}[ht!]
\centering
\includegraphics[width=11cm,angle=90]{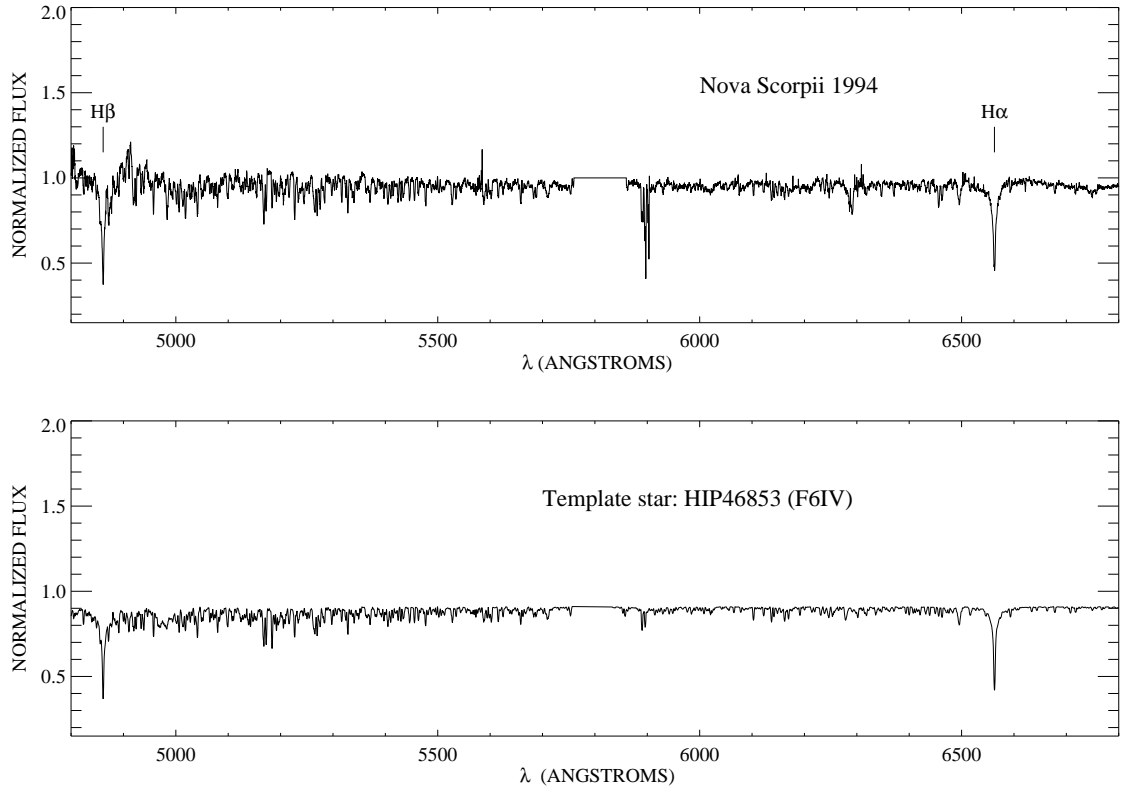}
\caption{\footnotesize{Observed spectrum of the secondary star of \mbox{Nova Scorpii
1994} (top) and of a properly broadened template (HIP 46583, bottom).}} 
\label{fig2}
\end{figure*}

\subsection{Secondary spectrum}

The individual UVES spectra were corrected for radial velocity
and combined in order to improve the signal-to-noise ratio. After
binning in wavelength with steps of 0.3 {\AA} the final UVES spectrum had a
signal-to-noise ratio of 120--150 in the continuum in the red part of
the spectrum. This spectrum is displayed in Fig.~\ref{fig2}.
Following Marsh et al. (1994), we computed the optimal $v~\sin~i$ by
subtracting broadened versions of a template star (HIP 46583), in
steps of 1 \kmso, and minimizing the residual. We used an spherical
rotational profile with linearized limb-darkening $\epsilon = 0.55$
(Al-Naimiy 1978). The best fit corresponds to a $v \sin
i=87^{+8}_{-4}$ \kmso, where the errors have been derived by assuming
extreme cases for $\epsilon = 0-1$. These values are consistent with previous studies
($v\sin i=86^{+3.3}_{-3.6}$ \kmso, Shahbaz et al.\ 1999). Our 
derived rotational velocity, combined with our value of the velocity
amplitude, $K_2$, implies a binary mass ratio $q=0.329\pm0.047$.
Shahbaz (2003) might have determined a more accurate value of the mass
ratio, $q=0.419\pm0.028$, by comparing synthetic spectra, which
incorporate the secondary's Roche geometry, with an intermediate 
resolution spectrum of \mbox{Nova Sco 94}.

\subsection{Diffuse interstellar bands}

We obtained the spectrum of the interstellar medium (ISM) from our own
observations. Due to the radial velocity of each individual
exposure, in the average spectrum of all the individual spectra, the
interstellar medium features are smoothed out. We subtracted the
stellar features in each individual spectrum and combined them to
obtain a spectrum that contains the ISM features. We tentatively
identified in this spectrum 31 diffuse interstellar bands with
equivalent widths typically in the range 10--65 m{\AA}, although there
are strong ISM features at 6060, 6202, 6283 and 6613 {\AA} with equivalent
widths of 148, 207, 555 and 175 m{\AA}, respectively (Galazutdinov et
al.\ 2000). In addition, the Na doublet at 5889 and 5895 {\AA} show
equivalent width of 1.124 and 1.006 {\AA}. We took care 
that these bands did not affect the lines selected for the chemical
analysis described in the next section. 

\section{Chemical Analysis}

\subsection{Stellar Parameters\label{stam}}

In the normalized spectra of X-ray transients, although observed
at quiescence, the stellar lines of the secondary star may be
veiled by the flux coming from the accretion disc. This effect is more
relevant in secondary stars of low luminosity. In the optical
range, this veiling appears to drop toward longer wavelengths (e.g.
Marsh et al.\ 1994). In our analysis, we tried to
infer the stellar atmospheric parameters of the secondary star, using
a technique which compares a grid of synthetic spectra with the
observed VLT/UVES spectrum, via a $\chi2$-minimization procedure. This
procedure takes into account the effect that veiling causes on the
stellar lines and it has been applied in other X-ray binaries
(A0620--00, Gonz\'alez Hern\'andez et al. 2004; Centaurus X-4, 2005;
and XTE J1118+480, 2006). For Nova Sco 94, we have also considered
any possible veiling from the accretion disc. 

For this purpose, we firstly identified moderately 
strong and relatively unblended lines of the elements of interest in the
high resolution solar atlas of Kurucz et al. (1984). We compute
synthetic spectra with the local thermodynamic equilibrium (LTE) code
MOOG (Sneden 1973), adopting the atomic line data from the Vienna
Atomic Line Database (VALD, Piskunov et al. 1995) and using a grid of
LTE model atmospheres (Kurucz 1993). We adjusted the oscillator
strengths of the selected lines until 
reproducing both the solar atlas of Kurucz et al. (1984) with solar
abundances (Grevesse et al.\ 1996) and the spectrum of Procyon with
its derived abundances (Andrievsky et al.\ 1995). The changes we
applied to the $\log gf$ values taken from VALD database are $\Delta\log
gf \lesssim 0.2$ dex. However, in some cases, we modified the $\log gf$
values even in the range 0.2--0.4 dex for some lines as for instance,
the \ion{O}{i} $\lambda6156$--8 {\AA} lines and the \ion{Mg}{i}
$\lambda5528$ {\AA} and $\lambda6318$--9 {\AA} lines (see Tables~\ref{tbl2}
and~\ref{tbl3}).  

\begin{table}
\centering
\caption[]{Ranges and steps of model parameters}
\begin{tabular}{lcc}
\noalign{\smallskip}
\noalign{\smallskip}
\noalign{\smallskip}
\hline
\hline
\noalign{\smallskip}
Parameter & Range & Step\\
\noalign{\smallskip}
\hline
\noalign{\smallskip}
$T_{\mathrm{eff}}$  & $5500 \rightarrow 7500$ K & 100 K\\ 
$\log (g/{\rm cm~s}^2)$  & $3. \rightarrow 5$  & 0.1\\ 
$\mathrm{[Fe/H]}$ & $-0.5 \rightarrow 1$  & 0.05\\ 
$f_{4500}$ &  $0 \rightarrow 0.6$  & 0.05\\ 
$m_0$  & $0 \rightarrow -0.00027$ & -0.00003\\ 
\noalign{\smallskip}
\hline	     
\hline
\end{tabular}
%
%
\label{tbl1}
\end{table}

\begin{figure*}[ht!]
\centering
\includegraphics[width=12cm,angle=0]{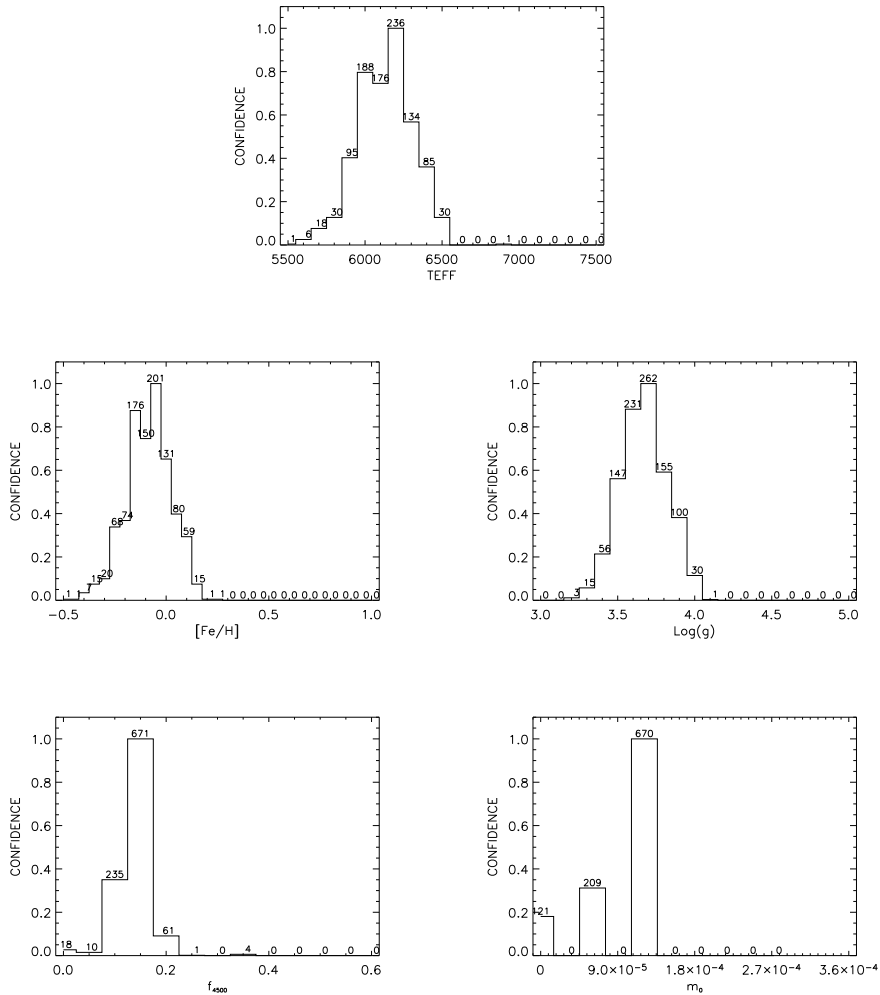}
\caption{\footnotesize{Distributions obtained for each parameter using
Monte Carlo simulations. The labels at the top of each bin indicate the
number of simulations consistent with the bin value. The total number
of simulations was 1000.}}
\label{fig3}
\end{figure*}

We selected six spectral features containing in total 42 lines of
\ion{Fe}{i} with excitation potentials between 1 and 5 eV. We
generated a grid of synthetic spectra for these features varying as
free parameters, the star effective tem\-pe\-ra\-tu\-re
($T_{\mathrm{eff}}$), the surface gravity ($\log g$), the metallicity
([Fe/H]), and the veiling from the accretion disc, which was assumed to be
a linear function of wavelength, and thus charaterized by two
parameters, its value at 4500 {\AA}, $f_{4500}=F^{4500}_{\rm
disc}/F^{4500}_{\rm sec}$, and the slope, $m_0$. Note that the total
flux is defined as $F_{\rm total} = F_{\rm disc}+F_{\rm  sec}$, where
$F_{\rm disc}$ and $F_{\rm sec}$ are the flux contributions of the
disc and the continuum of the secondary star, respectively. The
microturbulence, $\xi$, was fixed in each atmospheric model according to the
calibration as a function of effective temperature and surface gravity
reported by Allende Prieto et al. (2004). Such calibration has been
derived for stars of the solar neighbourhood with similar metallicity
as the secondary star in \mbox{Nova Sco 94}. For the stellar
parameters of the secondary star, determined below, this calibration
provides a microturbulence of $\xi=1.9$ \kmso.

The five free parameters were varying using the ranges and steps given in 
Table~\ref{tbl1}. In Fig.~\ref{fig3} we show the resulting 
histograms of these parameters corresponding to 1000 realizations of
observed spectrum. We have not found any clear evidence of veiling
from the accretion disc in the spectral region analysed
($\lambda\lambda$5400--6800 {\AA}), as expected for the spectral type
of the secondary star. The comparison of this grid, using
a bootstrap Monte-Carlo method, gives the most likely values
$T_{\mathrm{eff}} = 6100 \pm 200$ K, $\log (g/{\rm cm~s}^2) = 3.7 \pm
0.2$, $\mathrm{[Fe/H]} = -0.1 \pm 0.1$, $f_{4500} = 0.15\pm0.05$, and
$m_0 = -0.00012\pm0.00003$. This metallicity determination is
consistent with the values reported in the literature
($\mathrm{[Fe/H]} = 0.1 \pm 0.2$, Israelian et al. 1999;
$\mathrm{[Fe/H]} = -0.1 \pm 0.15$, Buxton \& Vennes 2001), whereas 
the effective temperature is slightly lower than the values obtained 
by these authors ($T_{\mathrm{eff}} = 6400 \pm 250$ K,
Israelian et al. 1999; $T_{\mathrm{eff}} = 6500 \pm 50$ K,
Buxton \& Vennes 2001), although consistent within the error bars.
We should remark that these authors used a different approach to
derive the stellar pa\-ra\-me\-ters. Israelian et al. (1999) adopted the
stellar parameters from previous calibrations of the spectral type
and luminosity classification of the secondary star as a F6III/F7IV
star. On the other hand, Buxton \& Vennes (2001) determined the
effective temperature by fitting a lower quality spectrum, in
particular, the wings of H$_\alpha$ profile.

\begin{table*}
\caption{Adopted $\log gf$ values of important lines in the VLT/UVES
spectrum. This line list do not included the weaker lines which these
lines are blended with. The best fit abundance computed in LTE,
[X/H], is also given for each feature.}    
\centering          
\renewcommand{\footnoterule}{}  
\begin{tabular}{lccrrlccrr}     
\noalign{\smallskip}
\noalign{\smallskip}
\noalign{\smallskip}
\hline
\hline       
\noalign{\smallskip}
Species & $\lambda$ ({\AA})& $\chi$ (eV) & $\log gf$ & [X/H] & Species &
$\lambda$ ({\AA})& $\chi$ (eV) & $\log gf$ & [X/H] \\ 
\noalign{\smallskip}
\hline
\noalign{\smallskip}
{Fe}~\scriptsize{I} & 5462.970 & 4.473 & -0.15 &   --  & {S}~\scriptsize{I}  & 6748.682 & 7.868 & -0.80 &   --  \\
{Fe}~\scriptsize{I} & 5463.276 & 4.434 &  0.11 & -0.20 & {S}~\scriptsize{I}  & 6748.837 & 7.868 & -0.60 &  0.70 \\
{Fe}~\scriptsize{I} & 5466.396 & 4.371 & -0.63 &   --  & {S}~\scriptsize{I}  & 6757.007 & 7.870 & -0.90 &   --  \\
{Fe}~\scriptsize{II}& 5534.847 & 3.245 & -2.76 &   --  & {S}~\scriptsize{I}  & 6757.171 & 7.870 & -0.31 &  0.55 \\
{Fe}~\scriptsize{I} & 5535.418 & 4.186 & -0.96 &  0.15 & {Si}~\scriptsize{I} & 5675.417 & 5.619 & -1.23 &  0.25 \\
{Fe}~\scriptsize{I} & 5554.895 & 4.548 & -0.39 & -0.20 & {Si}~\scriptsize{I} & 5675.756 & 5.619 & -2.08 &   --  \\
{Fe}~\scriptsize{I} & 5615.644 & 3.332 &  0.05 & -0.20 & {Si}~\scriptsize{I} & 5684.484 & 4.954 & -1.70 &  0.60 \\
{Fe}~\scriptsize{I} & 6191.558 & 2.433 & -1.52 & -0.10 & {Si}~\scriptsize{I} & 6112.928 & 5.616 & -2.25 &  0.75 \\
{Fe}~\scriptsize{I} & 6230.723 & 2.559 & -1.28 & -0.10 & {Si}~\scriptsize{I} & 6113.120 & 5.616 & -2.58 &   --  \\
{Fe}~\scriptsize{I} & 6393.610 & 2.433 & -1.51 & -0.10 & {Si}~\scriptsize{I} & 6131.565 & 5.616 & -1.70 &  0.25 \\
{Fe}~\scriptsize{I} & 6400.001 & 3.602 & -0.41 & -0.20 & {Si}~\scriptsize{I} & 6131.855 & 5.616 & -1.70 &   --  \\
{Fe}~\scriptsize{I} & 6592.914 & 2.727 & -1.67 & -0.05 & {Si}~\scriptsize{I} & 6142.204 & 5.619 & -2.50 &   --  \\
{Ca}~\scriptsize{I} & 6102.723 & 1.879 & -0.67 & -0.25 & {Si}~\scriptsize{I} & 6142.483 & 5.619 & -1.55 &  0.50 \\
{Ca}~\scriptsize{I} & 6122.220 & 1.886 & -0.01 &  0.10 & {Si}~\scriptsize{I} & 6145.016 & 5.616 & -1.42 &   --  \\
{Ca}~\scriptsize{I} & 6161.297 & 2.523 & -1.35 & -0.25 & {Si}~\scriptsize{I} & 6155.150 & 5.619 & -0.82 &  0.35 \\
{Ca}~\scriptsize{I} & 6162.180 & 1.899 &  0.37 &   --  & {Si}~\scriptsize{I} & 6155.693 & 5.619 & -2.22 &   --  \\
{Ca}~\scriptsize{I} & 6169.042 & 2.523 & -0.78 &  0.05 & {Si}~\scriptsize{I} & 6237.319 & 5.614 & -1.15 &  0.70 \\
{Ca}~\scriptsize{I} & 6169.563 & 2.526 & -0.45 &   --  & {Si}~\scriptsize{I} & 6238.286 & 5.082 & -2.38 &   --  \\
{Ca}~\scriptsize{I} & 6439.075 & 2.526 &  0.39 &  0.00 & {Si}~\scriptsize{I} & 6331.956 & 5.082 & -2.40 &  0.85 \\
{Ca}~\scriptsize{I} & 6471.662 & 2.526 & -0.71 &  0.15 & {Si}~\scriptsize{I} & 6433.457 & 5.964 & -1.62 &  0.90 \\
{Ca}~\scriptsize{I} & 6475.236 & 4.131 & -1.22 & -0.35 & {Si}~\scriptsize{I} & 6721.848 & 5.863 & -1.17 &  0.70 \\
{Ca}~\scriptsize{I} & 6717.685 & 2.709 & -0.62 &   --  & {Ti}~\scriptsize{I} & 5953.160 & 1.887 & -0.52 &  0.20 \\
{Mg}~\scriptsize{I} & 5167.321 & 2.709 & -1.03 &  0.30 & {Na}~\scriptsize{I} & 5682.633 & 2.102 & -0.69 &  0.70 \\
{Mg}~\scriptsize{I} & 5172.684 & 2.712 & -0.40 &  0.00 & {Na}~\scriptsize{I} & 5688.207 & 2.104 & -0.25 &  0.55 \\
{Mg}~\scriptsize{I} & 5183.604 & 2.717 & -0.18 &  0.00 & {Na}~\scriptsize{I} & 6154.226 & 2.102 & -1.68 &  0.05 \\
{Mg}~\scriptsize{I} & 5528.405 & 4.346 & -0.62 &  0.60 & {Na}~\scriptsize{I} & 6159.482 & 2.104 & -1.36 & -0.05 \\
{Mg}~\scriptsize{I} & 6318.717 & 5.108 & -1.99 &  0.30 & {Al}~\scriptsize{I} & 6696.023 & 3.143 & -1.83 &  0.05 \\
{Mg}~\scriptsize{I} & 6319.237 & 5.108 & -2.25 &   --  & {Al}~\scriptsize{I} & 6696.185 & 4.022 & -1.83 &   --  \\
{Mg}~\scriptsize{I} & 6319.495 & 5.108 & -2.43 &   --  & {Al}~\scriptsize{I} & 6696.788 & 4.022 & -1.85 &   --  \\
{O}~\scriptsize{I}  & 6155.966 &10.740 & -1.01 &   --  & {Al}~\scriptsize{I} & 6698.673 & 3.143 & -1.87 &   --  \\
{O}~\scriptsize{I}  & 6156.756 &10.741 & -0.89 &   --  & {Ni}~\scriptsize{I} & 5475.425 & 3.833 & -1.95 &   --  \\
{O}~\scriptsize{I}  & 6156.776 &10.741 & -0.69 &  0.90 & {Ni}~\scriptsize{I} & 5476.900 & 1.826 & -0.89 &  0.35 \\
{O}~\scriptsize{I}  & 6158.176 &10.741 & -1.00 &   --  & {Ni}~\scriptsize{I} & 5625.312 & 4.089 & -0.70 &  0.10 \\
{O}~\scriptsize{I}  & 6158.186 &10.741 & -0.41 &   --  & {Ni}~\scriptsize{I} & 5628.335 & 4.089 & -1.27 &  0.00 \\
{O}~\scriptsize{I}  & 6300.230 & 0.000 & -9.81 &   --  & {Ni}~\scriptsize{I} & 6116.199 & 4.089 & -0.68 &  0.45 \\
{S}~\scriptsize{I}  & 6743.440 & 7.866 & -1.27 &   --  & {Ni}~\scriptsize{I} & 6116.199 & 4.266 & -0.82 &   --  \\
{S}~\scriptsize{I}  & 6743.531 & 7.866 & -0.92 &  0.50 & {Ni}~\scriptsize{I} & 6118.094 & 4.088 & -2.32 &   --  \\
{S}~\scriptsize{I}  & 6743.640 & 7.866 & -1.03 &   --  & {Ni}~\scriptsize{I} & 6191.171 & 1.676 & -2.55 & -0.10 \\
{S}~\scriptsize{I}  & 6748.573 & 7.868 & -1.39 &   --  &  &  &  &  &  \\
\noalign{\smallskip}
\hline
\noalign{\smallskip}
\noalign{\smallskip}
\noalign{\smallskip}
\end{tabular}
\label{tbl2}
\end{table*}

\begin{table*}
\caption{Same as Table~\ref{tbl2} but for the important lines in the
Keck~I/HIRES spectrum.}   
\centering          
\renewcommand{\footnoterule}{}  
\begin{tabular}{lccrrlccrr}     
\noalign{\smallskip}
\noalign{\smallskip}
\noalign{\smallskip}
\hline
\hline       
\noalign{\smallskip}
Species & $\lambda$ ({\AA})&
$\chi$ (eV) & $\log gf$ & [X/H] & Species & $\lambda$ ({\AA})&
$\chi$ (eV) & $\log gf$ & [X/H] \\
\noalign{\smallskip}
\hline
\noalign{\smallskip}
{Mg}~\scriptsize{I} & 7875.434 & 5.932 & -2.13 &   --  & {O}~\scriptsize{I}  & 7775.400 & 9.110 &  0.10 &   --  \\
{Mg}~\scriptsize{II}& 7877.054 & 9.996 &  0.39 &  0.75 & {O}~\scriptsize{I}  & 8446.249 & 9.521 & -0.46 &   --  \\
{Mg}~\scriptsize{I} & 7877.478 & 5.932 & -1.95 &   --  & {O}~\scriptsize{I}  & 8446.359 & 9.521 &  0.23 &  1.80 \\
{Mg}~\scriptsize{I} & 7881.669 & 5.933 & -1.51 &  0.70 & {O}~\scriptsize{I}  & 8446.759 & 9.521 &  0.01 &   --  \\
{Mg}~\scriptsize{I} & 8712.676 & 5.932 & -1.67 &  --   & {S}~\scriptsize{I}  & 8693.137 & 7.870 & -1.37 &   --  \\
{Mg}~\scriptsize{I} & 8712.689 & 5.932 & -1.37 &  0.60 & {S}~\scriptsize{I}  & 8693.931 & 7.870 & -0.51 &  0.90 \\
{Mg}~\scriptsize{I} & 8717.803 & 5.933 & -2.85 &   --  & {S}~\scriptsize{I}  & 8694.626 & 7.870 &  0.08 &   --  \\
{Mg}~\scriptsize{I} & 8717.816 & 5.933 & -1.85 &   --  & {Si}~\scriptsize{I} & 8646.371 & 6.206 & -1.94 &   --  \\
{Mg}~\scriptsize{I} & 8717.825 & 5.933 & -0.93 &  0.65 & {Si}~\scriptsize{I} & 8648.465 & 6.206 & -0.30 &  0.55 \\
{Mg}~\scriptsize{I} & 8736.006 & 5.946 & -1.93 &   --  & {Si}~\scriptsize{I} & 8728.010 & 6.181 & -0.61 &  0.60 \\
{Mg}~\scriptsize{I} & 8736.019 & 5.946 & -0.69 &  0.65 & {Si}~\scriptsize{I} & 8728.594 & 6.181 & -1.72 &   --  \\
{Mg}~\scriptsize{I} & 8736.019 & 5.946 & -1.97 &   --  & {Si}~\scriptsize{I} & 8729.282 & 6.181 & -2.30 &   --  \\
{Mg}~\scriptsize{I} & 8736.029 & 5.946 & -1.02 &   --  & {Ti}~\scriptsize{I} & 8434.957 & 0.848 & -0.88 &  0.35 \\
{O}~\scriptsize{I}  & 7771.960 & 9.110 &  0.45 &  2.00 & {Ti}~\scriptsize{I} & 8435.652 & 0.836 & -1.02 &   --  \\
{O}~\scriptsize{I}  & 7774.180 & 9.110 &  0.31 &   --  & {Ti}~\scriptsize{I} & 8438.923 & 2.256 & -0.79 &   --  \\
\noalign{\smallskip}
\hline
\noalign{\smallskip}
\noalign{\smallskip}
\noalign{\smallskip}
\end{tabular}
\label{tbl3}
\end{table*}

\begin{figure*}[ht!]
\centering
\includegraphics[width=11cm,angle=90]{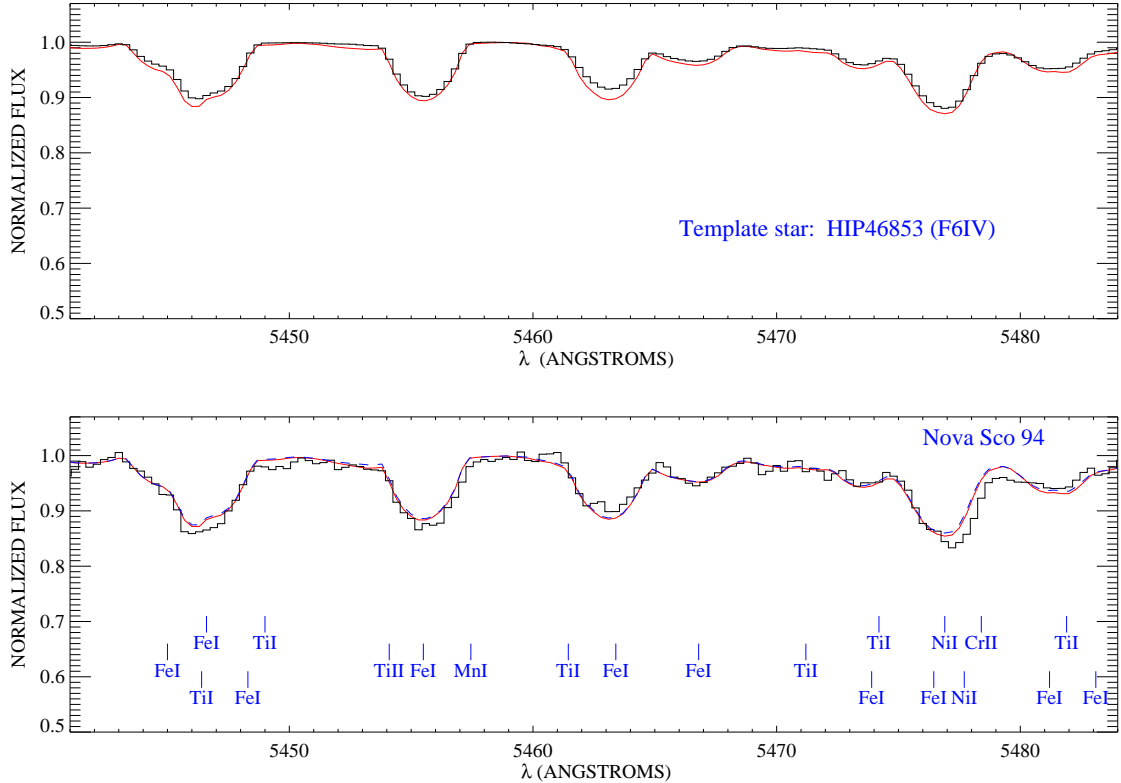}
\caption{\footnotesize{Best synthetic spectral fits to the UVES spectrum of the
secondary star in the \mbox{Nova Scorpii 1994} system (bottom
panel) and the same for a template star (properly broadened) shown
for comparison (top panel). Synthetic spectra are computed for
solar abundance ratios ($\mathrm{[E/Fe]}=0$ with
$\mathrm{[Fe/H]}=-0.1$, dashed line) and best fit abundance (solid
line).}}  
\label{fig4}
\end{figure*}

\begin{table}
\caption{Chemical abundances of the secondary star in \mbox{Nova
Scorpii 1994}}   
\centering          
\renewcommand{\footnoterule}{}  
\begin{tabular}{lccccccc}     
\noalign{\smallskip}
\noalign{\smallskip}
\noalign{\smallskip}
\hline
\hline
Element & $[{\rm X}/{\rm
H}]_{\rm LTE}^{\rm a,b}$ & $\sigma$ & $\Delta_\sigma^{\rm c}$ &
$\Delta_{T_{\rm eff}}$ & $\Delta_{\log g}$ & $\Delta_{\rm tot}$ &  $n_{\rm lines}^{\rm d}$\\   
\noalign{\smallskip}
\hline     
\noalign{\smallskip}
O$^{\rm e}$  &  0.91 & 0.07 & 0.04 & 0     &  0.07 & 0.09 & 3  \\
Na           &  0.31 & 0.37 & 0.18 & 0.19  &  0    & 0.26 & 4  \\
Mg           &  0.48 & 0.29 & 0.09 & 0.10  & -0.03 & 0.15 & 10 \\
Al           &  0.05 & 0.15 & 0.15 & 0.1   &  0    & 0.18 & 1  \\
Si           &  0.58 & 0.21 & 0.06 & 0.05  &  0.01 & 0.08 & 12 \\
S            &  0.66 & 0.18 & 0.09 & -0.04 &  0.06 & 0.12 & 4  \\
Ca           & -0.02 & 0.17 & 0.07 & 0.12  &  0.01 & 0.14 & 7  \\
Ti           &  0.27 & 0.11 & 0.07 & 0.2   &  0.05 & 0.22 & 2  \\
Fe           & -0.11 & 0.11 & 0.03 & 0.09  &  0.01 & 0.10 & 9  \\
Ni           &  0    & 0.10 & 0.06 & 0.2   &  0.05 & 0.21 & 5  \\
Li$^{\rm f}$ &  2.13 & 0.15 & 0.15 & 0.15  &  0    & 0.21 & 1  \\ 
\noalign{\smallskip}
\hline
\noalign{\smallskip}
\noalign{\smallskip}
\noalign{\smallskip}
\end{tabular}
\begin{minipage}[t]{\columnwidth}

$^{\rm a}\:$ The solar element abundances were adopted from Grevesse
et al. (1996) and Ecuvillon et al. (2006):
$\log \epsilon(\mathrm{O })_{\odot}=8.74$, 
$\log \epsilon(\mathrm{Na})_{\odot}=6.33$,
$\log \epsilon(\mathrm{Mg})_{\odot}=7.58$,  
$\log \epsilon(\mathrm{Al})_{\odot}=6.47$,  
$\log \epsilon(\mathrm{Si})_{\odot}=7.55$,  
$\log \epsilon(\mathrm{S })_{\odot}=7.21$,  
$\log \epsilon(\mathrm{Ca})_{\odot}=6.36$,  
$\log \epsilon(\mathrm{Ti})_{\odot}=4.99$,  
$\log \epsilon(\mathrm{Fe})_{\odot}=7.50$,  
$\log \epsilon(\mathrm{Ni})_{\odot}=6.25$

$^{\rm b}\:$Element abundances (calculated assuming LTE) are
$\mathrm{[X/H]}= \log [N(\mathrm{X})/N(\mathrm{H})]_{\rm star} - \log
[N(\mathrm{X})/N(\mathrm{H})]_{\rm Sun}$, where $N(\mathrm{X})$ is
the number density of atoms. Uncertainties, $\Delta {\rm tot}$, are at
1$\sigma$ level and take into account the error in the
stellar parameters and the dispersion of the measurements.

$^{\rm c}\:$The errors from the dispersion of the best
fits to different features, see text.

$^{\rm d}\:$Number of features analysed for each element.

$^{\rm e}\:$NLTE abundance of oxygen, see text.

$^{\rm f}\:$\mbox{Li} abundance is expressed as: $$\log
\epsilon(\mathrm{Li})_\mathrm{NLTE} = \log
[N(\mathrm{Li})/N(\mathrm{H})]_\mathrm{NLTE} + 12$$ 

\end{minipage}
\label{tbl4}
\end{table}

\subsection{Stellar abundances}

We inspected several spectral regions in both VLT/UVES and
Keck~I/HIRES spectra, searching for suitable lines for a detailed
chemical analysis. Using the derived stellar parameters, we firstly 
determined the Fe abundance from each individual feature in the
observed UVES spectrum (see Table~\ref{tbl2}). Since we knew of a
potential enhancement of $\alpha-$elements in the atmosphere of this
star (Israelian et al. 1999), we computed new LTE model atmospheres
with $\alpha-$elements enhanced by +0.4 dex with respect to solar
using the Linux version (Sbordone et al. 2004) of the ATLAS code
(Kurucz 1993). We used the new Opacity Distribution Functions (ODFs)
of Castelli \& Kurucz (2003) with the corresponding metallicity of the
secondary star. In Fig.~\ref{fig4} we show some of the spectral
regions analysed to obtain the Fe abundance. This figure also show the
best synthetic spectral fit to the observed spectrum of a template
star (HIP 46853 with \teffo$=6310$ K, \loggo$=3.91$ and [Fe/H]$=-0.08$
dex) using the stellar parameters and abundances determined by
Allende-Prieto et al. (2004). Thus, we only use as abundance
indicators those features which were well reproduced in the template
star. The chemical analysis is summarized in Table~\ref{tbl4}. For the
elements O, Mg, S, Si, Ti, previously studied by Israelian et al.
(1999), we selected the most suitable lines in the UVES and HIRES
spectra, whereas for other elements Na, Al, Ca, Ni, Fe and Li, we only
used spectral features in the VLT/UVES spectrum. In
Table~\ref{tbl4} we provide the average abundance of each element
extracted from the analysis of several features. The number of
features analysed for each element is also stated. Individual
element abundances derived from each feature are given in
Table~\ref{tbl2} and Table~\ref{tbl3}. The errors on the
element abundances show their sensitivity to the uncertainties on the
effective temperature, $\Delta_{T_{\mathrm{eff}}}$, surface gravity,
$\Delta_{\log g}$, and the dispersion of the measurements from
different spectral features, $\Delta_{\sigma}$. The errors
$\Delta_{\sigma}$ were estimated as $\Delta_{\sigma}
=\sigma/\sqrt{N}$, where $\sigma$ is the standard deviation of the $N$
measurements. The errors $\Delta_{T_{\mathrm{eff}}}$ and $\Delta_{\log
g}$ were determined as $\Delta_{T_{\rm
eff}}=(\sum_{i=1}^N\;\Delta_{T_{\rm eff},i})/N$ and $\Delta_{\log
g}=(\sum_{i=1}^N\;\Delta_{T_{\log g},i})/N$. The Al and Li abundance
were derived from one spectral line (see Fig.~\ref{fig8}) and the
error associated to the dispersion of the measurements, $\sigma$, was
assumed to be the average dispersion of Fe, Ca and Ni abundances, and
in this case, $\Delta_{\sigma}=\sigma$. The total error given in
Table~\ref{tbl4} was derived using the following expresion:
$\Delta_{\rm tot} = \sqrt{\Delta_{\sigma}^2 +
\Delta_{T_{\mathrm{eff}}}^2 + \Delta_{\log g}^2}$.  

\subsubsection{Oxygen}

\begin{figure*}[ht!]
\centering
\includegraphics[width=11cm,angle=90]{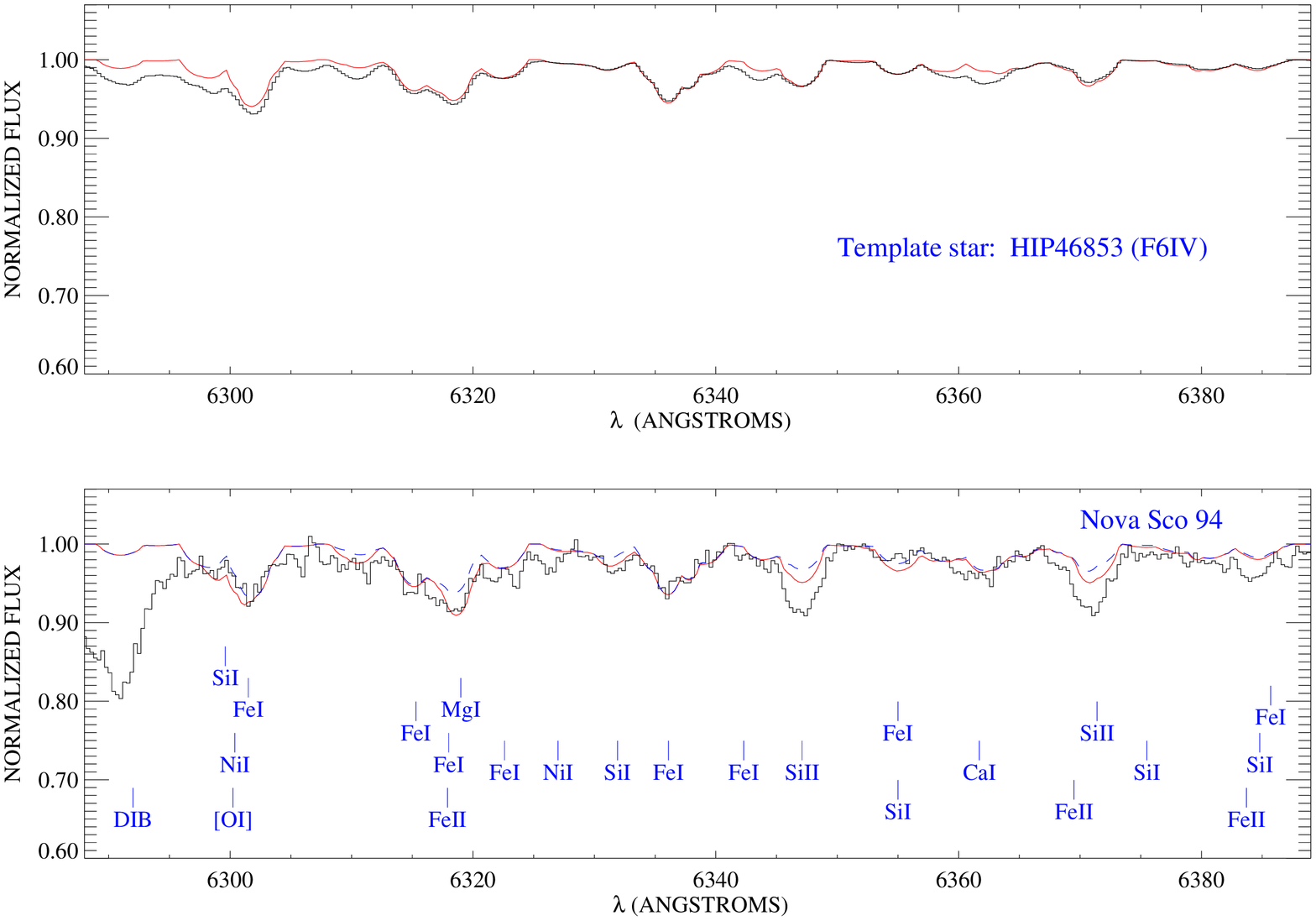}
\caption{\footnotesize{The same as in Fig.\ 4, but for the spectral range
$\lambda\lambda6275$--6400 {\AA}.}}  
\label{fig5}
\end{figure*}

The atomic data for \ion{O}{i} lines were adopted from
Ecuvillon et al. (2006) where oscillator strengths were slightly
modified in order to obtain a solar oxygen abundance of $\log
\epsilon(\mathrm{O})_{\odot}=8.74$. In Fig.~\ref{fig5} we display one
of the spectral regions analysed, where the forbidden line
[\ion{O}{i}] $\lambda6300.2$ {\AA} is present. Unfortunately, we found this
line inconvenient to provide a reliable abundance: it is very weak at
the effective temperature of the secondary star; it is blended with
stronger lines of \ion{Si}{i} $\lambda6299.6$ {\AA} and \ion{Fe}{i}
$\lambda6301.5$ due to high rotational broadening of the star; and the
strong narrow diffuse band (see the feature labelled DIB in
Fig.~\ref{fig5}) at 6283.8 {\AA} makes it difficult to place properly
the continuum. An oxygen abundance of $[{\rm O}/{\rm H}]\sim0.9$
(solid line in  Fig.~\ref{fig5}) seems not to fit at all the
[\ion{O}{i}] feature which may be better reproduced using a solar
abundance synthetic spectrum (dashed line in Fig.~\ref{fig5}). 

\begin{figure*}[ht!]
\centering
\includegraphics[width=6cm,angle=90]{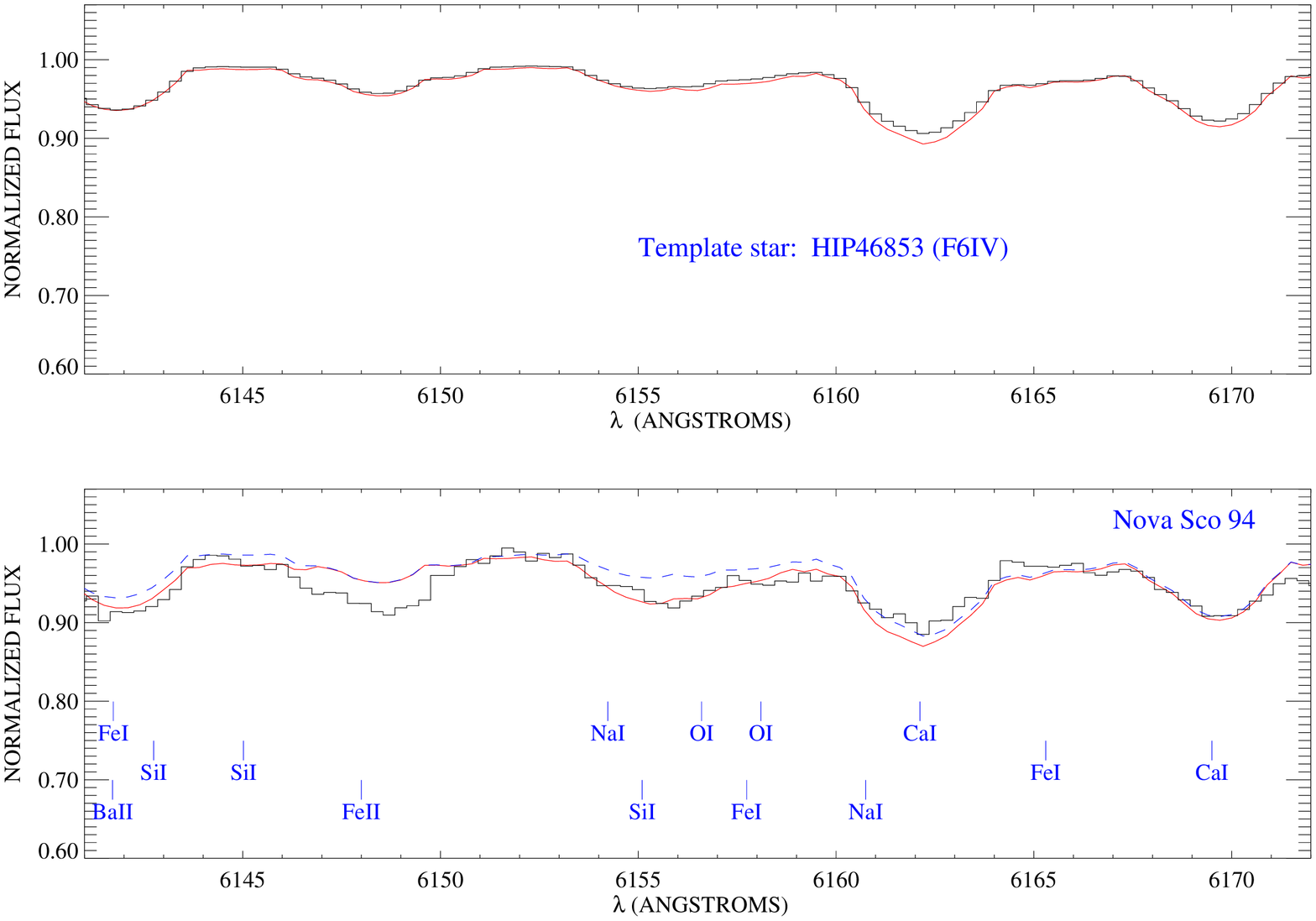}
\includegraphics[width=6cm,angle=90]{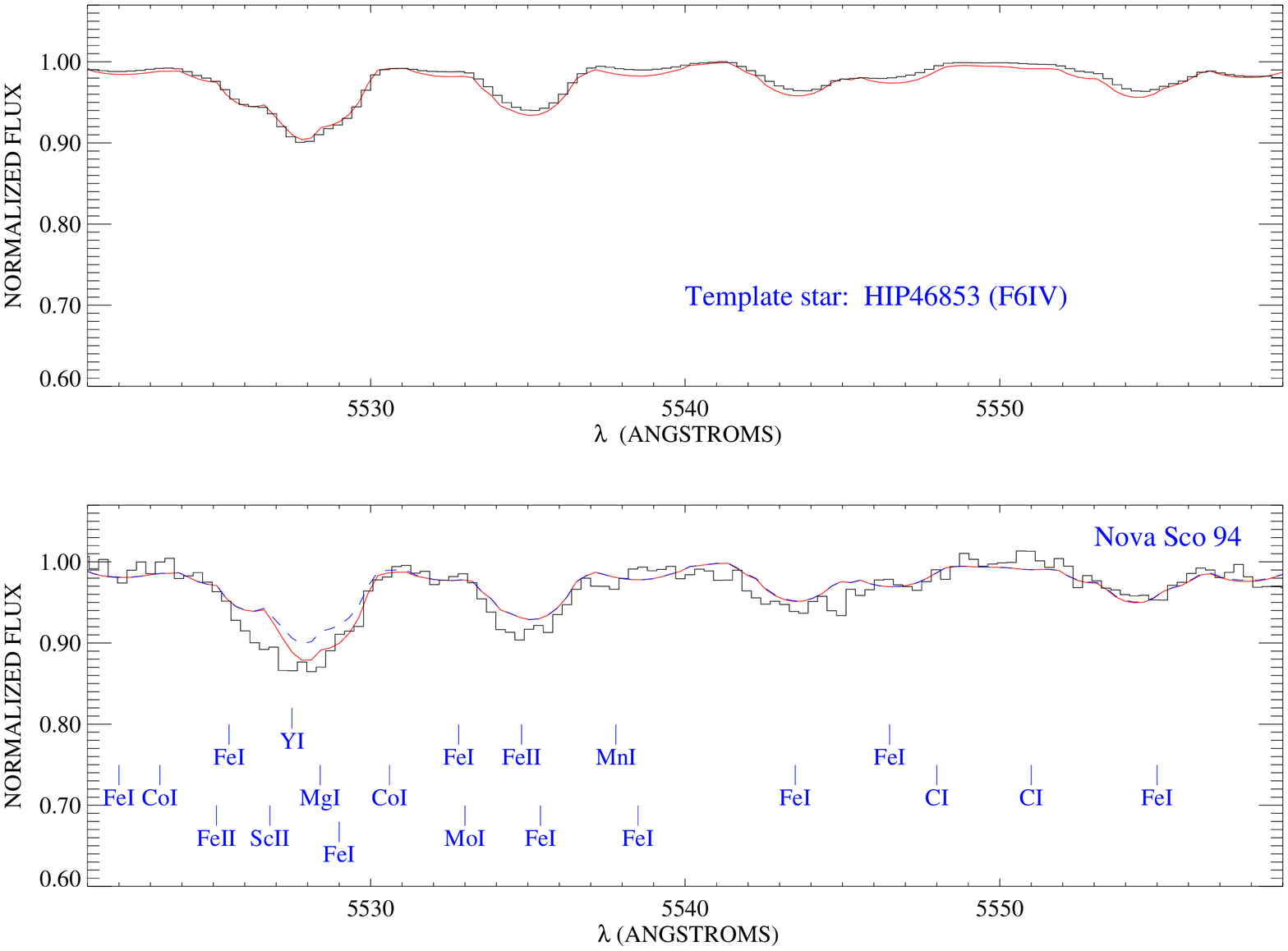}
\caption{\footnotesize{The same as in Fig.\ 4, but for the spectral
range $\lambda\lambda6141$--6172 {\AA} (left panels) and
$\lambda\lambda5521$--5559 {\AA} (right panels).}}   
\label{fig6}
\end{figure*}

Thus, we determined the oxygen abundance from three features
containing: \ion{O}{i} $\lambda6156$--8 {\AA} lines, which appear severely
blended with \ion{Si}{i}, \ion{Ca}{i} and \ion{Fe}{i} lines, and 
the  relatively isolated \ion{O}{i} $\lambda7771$--5 {\AA} and \ion{O}{i} 
8446 {\AA} triplets. In the left-bottom panel of Fig.~\ref{fig6}
we show the \ion{O}{i} $\lambda6156$--8 {\AA} lines in the VLT spectrum of the
target in comparison with several synthetic spectra in LTE. The best
fit in LTE gives an oxygen abundance of $\mathrm{[O/H]_{\rm
LTE}}\sim0.9$. While LTE appears to be practically valid for 
the \ion{O}{i} $\lambda6156$--8 {\AA} (Takeda et al.\ 2005), the \ion{O}{i} 
$\lambda7771$--5 {\AA} and \ion{O}{i} $\lambda8446$ {\AA} suffers appreciable non-LTE
effects (Ecuvillon et al.\ 2006). The NLTE
corrections\footnote{$\Delta_\mathrm{NLTE}=\log
\epsilon(\mathrm{X})_\mathrm{NLTE}-\log 
\epsilon(\mathrm{X})_\mathrm{LTE}$} were kindly provided by N.
Shchukina (private communication; Shchukina 1987; Ecuvillon et al.\
2006). For the stellar parameters and oxygen abundance of the
secondary star, NLTE corrections were estimated at -0.988 and 
-0.926 for the near-IR \ion{O}{i} $\lambda7771$--5 {\AA} and \ion{O}{i}
$\lambda8446$ {\AA} triplets, respectively. 

\begin{figure*}[ht!]
\centering
\includegraphics[width=6cm,angle=90]{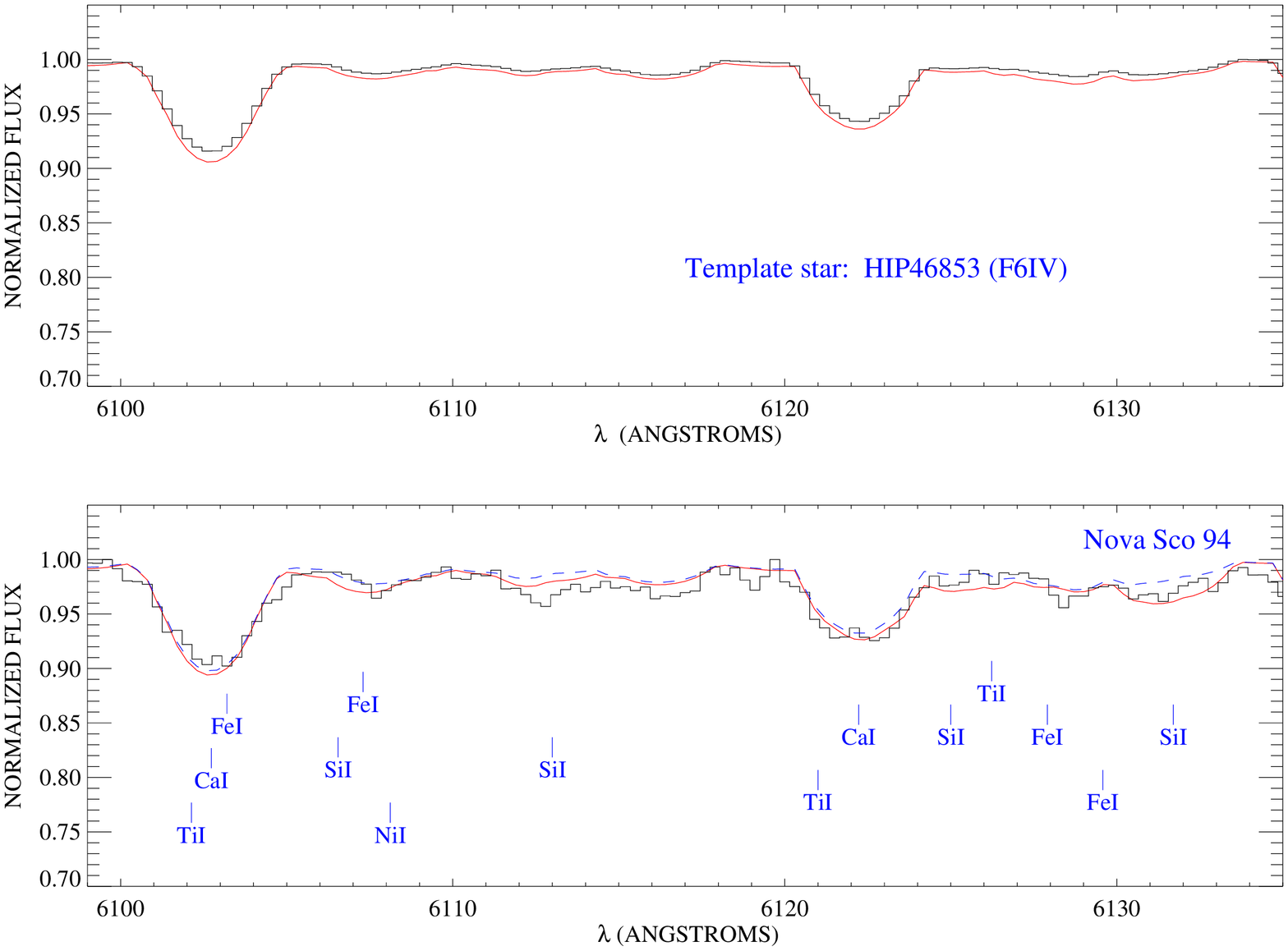}
\includegraphics[width=6cm,angle=90]{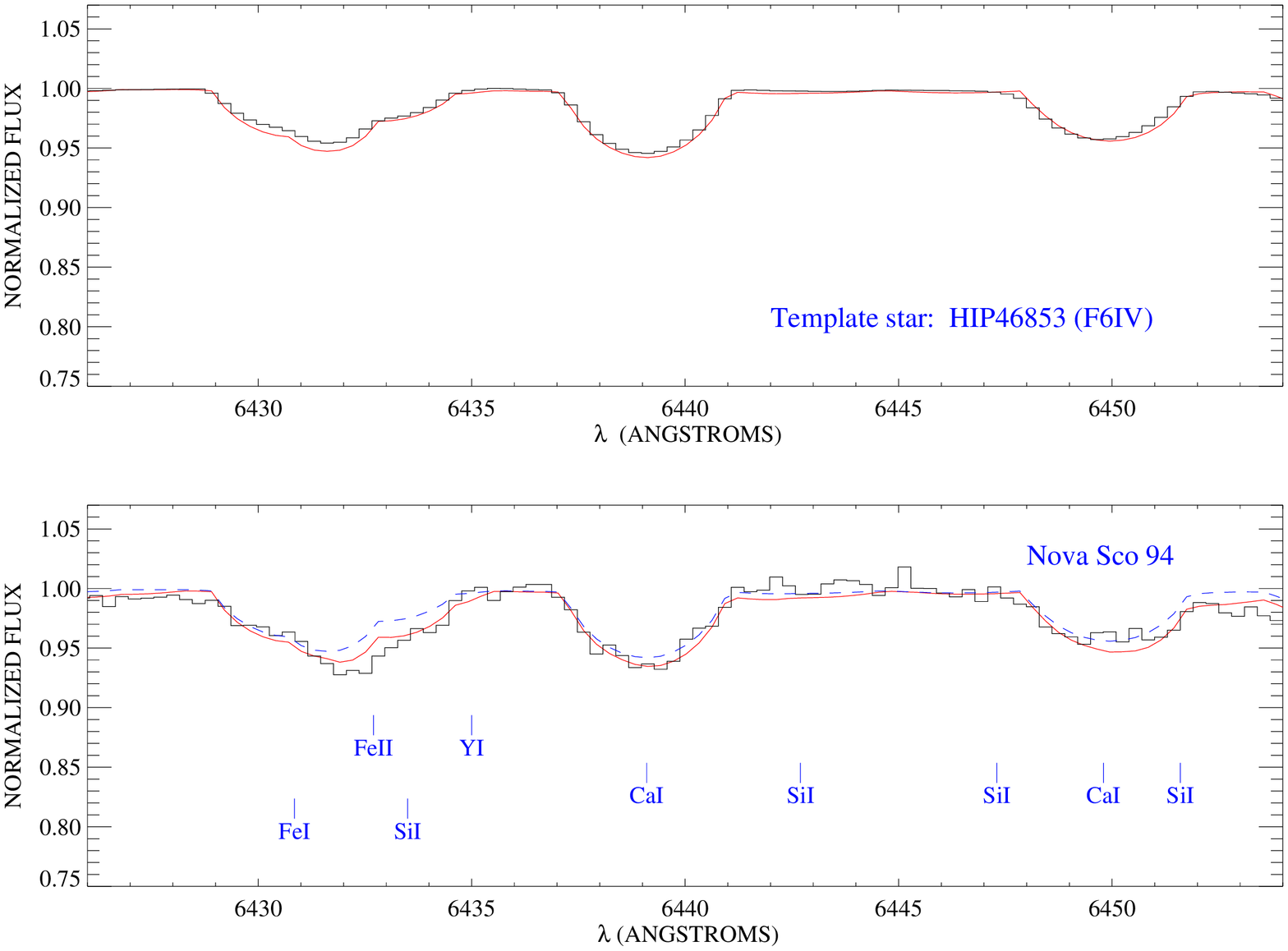}
\caption{\footnotesize{The same as in Fig.\ 4, but for the spectral
range $\lambda\lambda6099$--6135 {\AA} (left panels) and
$\lambda\lambda6426$--6454 {\AA} (right panels).}}   
\label{fig7}
\end{figure*}

\subsubsection{Magnesium\label{smg}}

The abundance of magnesium was determined from 5 \ion{Mg}{i}
spectral features in the optical VLT/UVES spectrum, including the
three features of the \ion{Mg}{i}b triplet at 5167--83 {\AA}, 4
\ion{Mg}{i} spectral features in the near-IR Keck~I/HIRES spectrum,
where \ion{Mg}{i} lines are isolated or weakly blended with other
lines, and a \ion{Mg}{ii} spectral feature at 7877 {\AA}. In
Fig.~\ref{fig5} and the right-bottom panel of Fig.~\ref{fig6}, we show
some of the \ion{Mg}{i} features included in the abundance analysis.
Note that the best fit synthetic spectra displayed in these figures
show the best fit abundance of each feature instead of the average
abundance of each element given in Table~\ref{tbl4}. We should remark
that the location of the continuum near the near-IR lines, where the
S/N$\sim50$, is less accurate, than that of the optical lines, where
the S/N$\sim 150$. Thus, we estimate an average uncertainty of $\sim
0.27$ dex in the Mg abundance due to the uncertainty on the continuum
location in the near-IR spectrum, whereas an uncertainty of
$\sim0.05-0.1$ dex is expected from the optical features. This might
explain the high dispersion of the Mg measurements given in
Table~\ref{tbl4}. On the other hand, the near-IR Mg features are less
sensitive ($\le0.1$ dex) to the errors on the stellar parametes than the
optical lines, specially the \ion{Mg}{i}b triplet whose errors related
to the uncertainties on the stellar parameters are
$\Delta_{T_{\mathrm{eff}}}\sim0.2-0.3$ dex, and $\Delta_{\log g}\sim
-0.1$ dex. More details on the possible uncertainties on the analysis of
\ion{Mg}{i}b lines are given in \S~\ref{smgib}. In addition, some of
the Mg lines analysed are known to exhibit non-LTE effects. NLTE
abundance corrections, $\Delta_\mathrm{NLTE}$, for \ion{Mg}{i} lines
have been studied by Zhao et al. (1998, 2000) in the Sun. Among the
lines we have analysed, they only reported corrections for the
\ion{Mg}{i} $\lambda5528$ {\AA} and the \ion{Mg}{i}b $\lambda5167-5183$
{\AA}, which are expected to show NLTE abundance corrections of roughly
$\sim +0.05$ dex in the Sun and, in general, less than 0.1 dex in
dwarfs and subdwarfs (Shimanskaya et al. 2000). On the contrary, Abia
\& Mashonkina (2004) estimated NLTE corrections of $-0.02 <
\Delta_\mathrm{NLTE} < -0.07$ in the Sun for the \ion{Mg}{ii} 
$\lambda7877$ {\AA}. Although these NLTE corrections are small, they
should be investigated for the case of the abundance and stellar
parameters of the secondary star in this system.

\subsubsection{Silicon}

The Si abundance was obtained from the 12 features with Si~I lines
often blended with other lines, some of them are displayed in
Figs.~\ref{fig7} and~\ref{fig8}. The dispersion of the abundance
measurements from different lines is 0.2 dex, although the average
abundance has a good accuracy due to the number of features analysed. As
shown in Fig.~\ref{fig5}, we found difficult to reproduce \ion{Si}{ii}
lines, surprisingly strong in comparison with \ion{Si}{i} lines, with
the average Si abundance, so that, we decided not to consider them as
abundance indicators. Shchukina \& Trujillo Bueno (2001) pointed out
that NLTE effects for Si (with an ionization potential of 8.15 eV,
slightly higher than that of Fe, 7.87 eV) would likely be similar to
those found for Fe, which suffers from over-ionization, in a
three-dimensional solar photospheric model. In addition, for species
where over-ionization is prominent, the NLTE effects normally increase
toward higher \teff and lower \logg (Asplund 2005).

Wedemeyer (2001) has performed a NLTE analysis of Si in the Sun and 
argued that most of the levels of \ion{Si}{i} show almost
negligible deviations from LTE, whereas \ion{Si}{ii} levels are
overpopulated with respect to LTE. In particular, this author predicts
NLTE corrections of -0.01 dex for one of the \ion{Si}{i} lines we have
analysed at 5684 {\AA}. For the \ion{Si}{ii} lines at 6347 {\AA} and
6371 {\AA} the NLTE corrections are -0.097 dex and -0.064 dex for the
Sun, respectively, and therefore 6 to 9 times higher for the \ion{Si}{ii}
lines than for the \ion{Si}{i} lines. These NLTE effects might be
significantly larger for the \ion{Si}{ii} lines in the secondary star,
giving rise to large NLTE corrections. The synthetic profile computed
in NLTE would make stronger \ion{Si}{ii} lines, thus providing a
better description of the observations. However, this conclusion must
be clarified in forthcoming investigations. Note that, in fact, the
synthetic spectral fits reproduce quite well the \ion{Si}{ii} features
in the template, but these features are much stronger in the spectrum
of the secondary star in \mbox{Nova Sco 94}.

\subsubsection{Sulphur}

\begin{figure*}
\centering
\includegraphics[width=11cm,angle=90]{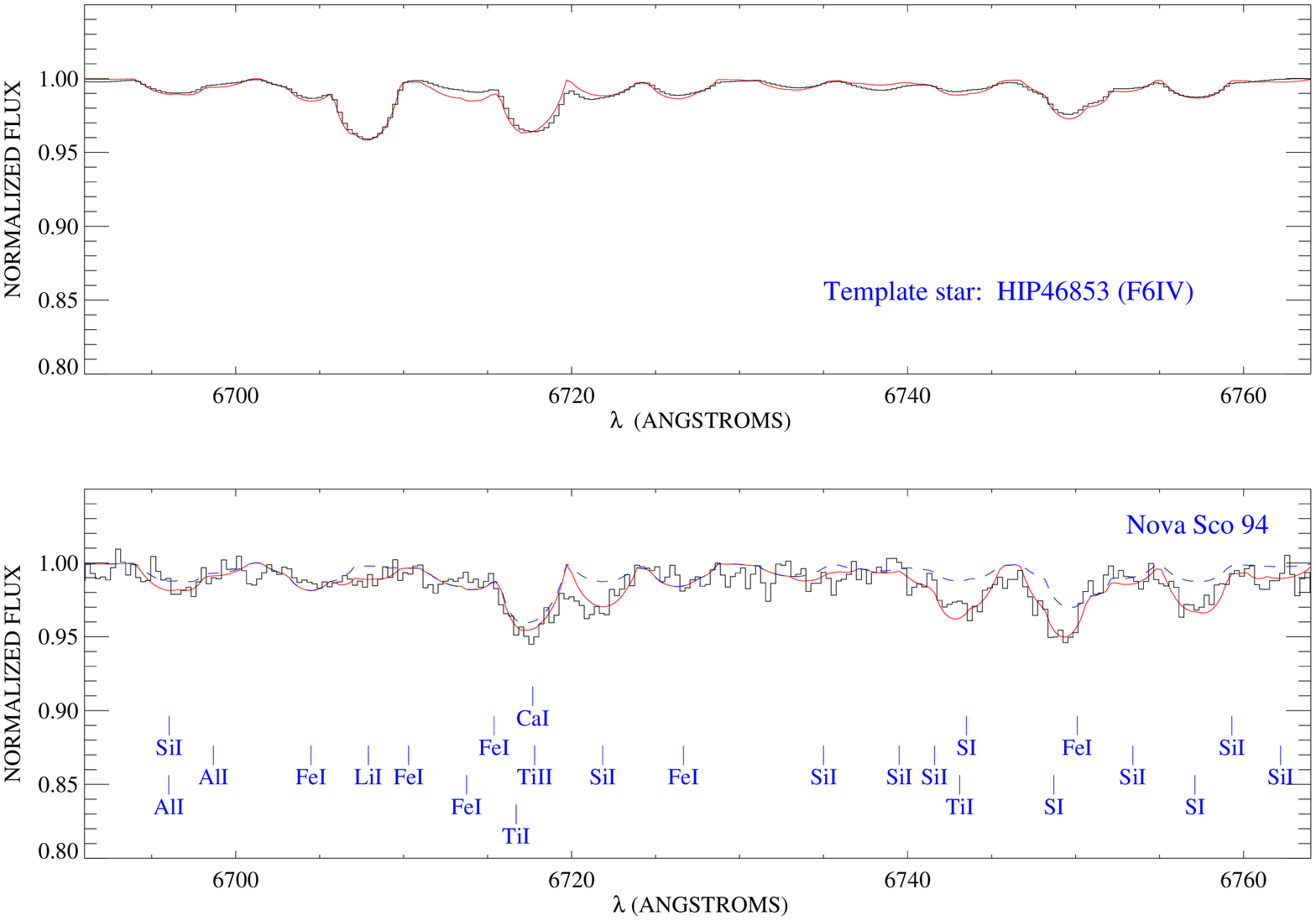}
\caption{\footnotesize{The same as in Fig.\ 4, but for the spectral
range $\lambda\lambda6690$--6760 {\AA}.}}  
\label{fig8}
\end{figure*}

In Fig.~\ref{fig8}, we show another spectral region with the strong
\ion{S}{i} $\lambda6743$--57 {\AA} lines. The overabundance of S is clearly seen
in the spectrum of \mbox{Nova Sco 94} when compared with the template.
These three strong and unblended lines of sulphur clearly confirm the
overabundance of sulphur in the secondary star of this system. We also
included one \ion{S}{i} feature at 8694 {\AA}, found in the
Keck spectrum, in the S abundance determination. Takeda-Hidai et al.
(2002) and Takeda et al. (2005) have reported negligible corrections
for \ion{S}{i} $\lambda8694$ {\AA} lines. For the stellar parameters,
metallicity and abundance ratio [S/Fe]$\sim +0.6$ dex of the
secondary star, NLTE corrections are comprised in the range $-0.04 <
\Delta_\mathrm{NLTE} < -0.08$ dex. These corrections were not
considered in the sulphur abundance given in Table~\ref{tbl4}, since
they are almost negligible compared to the dispersion of the LTE
abundances derived from different \ion{S}{i} lines. In addition, NLTE
corrections for the \ion{S}{i} $\lambda6743$--57 {\AA} lines are not available
in the literature.  

\subsubsection{Calcium}

Calcium is one the $\alpha$-elements that could not be measured in 
the analysis of the Keck spectrum reported in Israelian et al. (1999). 
The abundance of calcium was obtained from the analysis of 7 almost
unblended and strong spectral features in the VLT spectrum. Some of
these features are displayed in the left-bottom panel of
Figs.~\ref{fig6},~\ref{fig7} and~\ref{fig8}. NLTE abundance corrections have been
reported for the \ion{Ca}{i} $\lambda6102$, $\lambda6169$,
$\lambda6439$ and $\lambda6471$ {\AA} lines (Mashonkina et
al. 2007), being -0.05, -0.01, -0.13, -0.06 dex, respectively for a
model of \teffo$=6000$ K \loggo$=4$ dex and [Fe/H]$=0$ dex. This
results on an average correction $\Delta_\mathrm{NLTE} < -0.05$ dex
which does not have any significant impact on the Ca abundance.

\subsubsection{Titanium}

Israelian et al. (1999) could not measure a reliable abundance of Ti
using the Keck spectra since they did not find spectral features with
relatively unblended \ion{Ti}{i} lines. Unfortunately, the signal-to-noise
ratio in the spectral region from 5920 to 5960 {\AA} of the VLT/UVES
spectrum of \mbox{Nova Sco 94} where there are some Ti features is not
enough to provide reliable abundances. This spectral region is close
the edge of one of the arms of the UVES spectrograph. Thus, we finally
selected one of these features in the VLT spectrum and another
relatively unblended feature of Ti in the Keck
spectrum. We emphasize that the Ti abundance must be considered with
caution, although the UVES spectrum seems to indicate that Ti
abundance is not as enhanced as other $\alpha$-elements like O, Mg, Si and
S, but it shows an abundance almost consistent with solar if we take
into account the error bars (see Table~\ref{tbl4}).

\begin{figure*}
\centering
\includegraphics[width=11cm,angle=90]{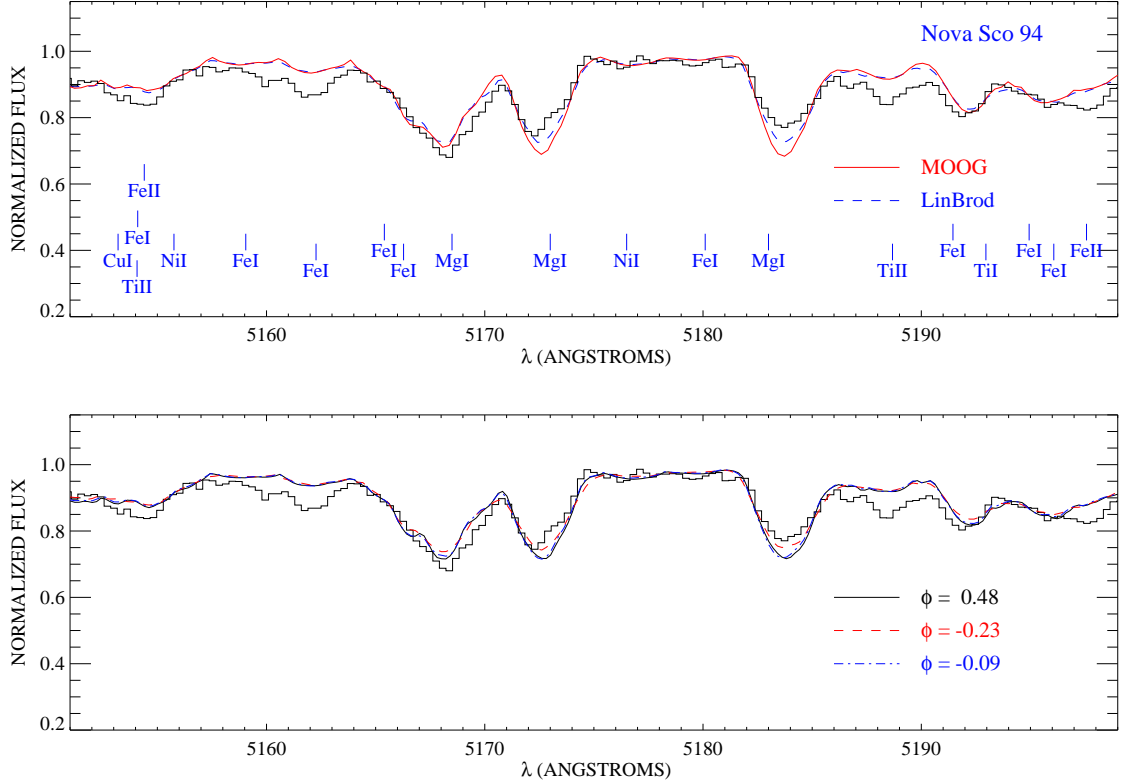}
\caption{\footnotesize{Upper panel: Observed UVES spectrum of the
secondary star in comparison with the average synthetic spectrum
computed with {\scshape LinBrod} (dashed line) for all orbital phases
and the synthetic spectrum computed with MOOG (solid line). An
abundance of [Mg/H]$\sim 0.3$ dex has been adopted for all synthetic
spectra. Bottom panel: Synthetic spectra computed with {\scshape
LinBrod} for different orbital phases.}}     
\label{fig9}
\end{figure*}

\subsubsection{The odd-Z elements: Na, Al}

The \ion{Al}{i} $\lambda6696$--8 {\AA} lines appear to be surprisingly weak
since one could expect to obtain an overabundance of aluminum
according to supernova yields (Umeda \& Nomoto, 2002, 2005; Tominaga,
Umeda \& Nomoto 2007; Maeda et al. 2002). Baum\"uller
\& Gehren (1996, 1997) show that this doublet is not sensitive to NLTE
conditions, with expected corrections $\Delta_\mathrm{NLTE} < +0.05$
for models with similar stellar parameters as the secondary star.

The sodium abundance was determined from 4 features, typically blended
with other lines (see Table~\ref{tbl4}). Its abundance is not as high
as expected from yields of SN models which suggest a similar behaviour
as aluminum. Baum\"uller et al. (1998) predict NLTE corrections
 $\Delta_\mathrm{NLTE} < -0.03$ and $\Delta_\mathrm{NLTE} < -0.05$ 
for the \ion{Al}{i} $\lambda5682$--8 and $\lambda6154$--60 lines,
respectively, which are not relevant in the abundance analysis.

\begin{figure*}
\centering
\includegraphics[width=11cm,angle=90]{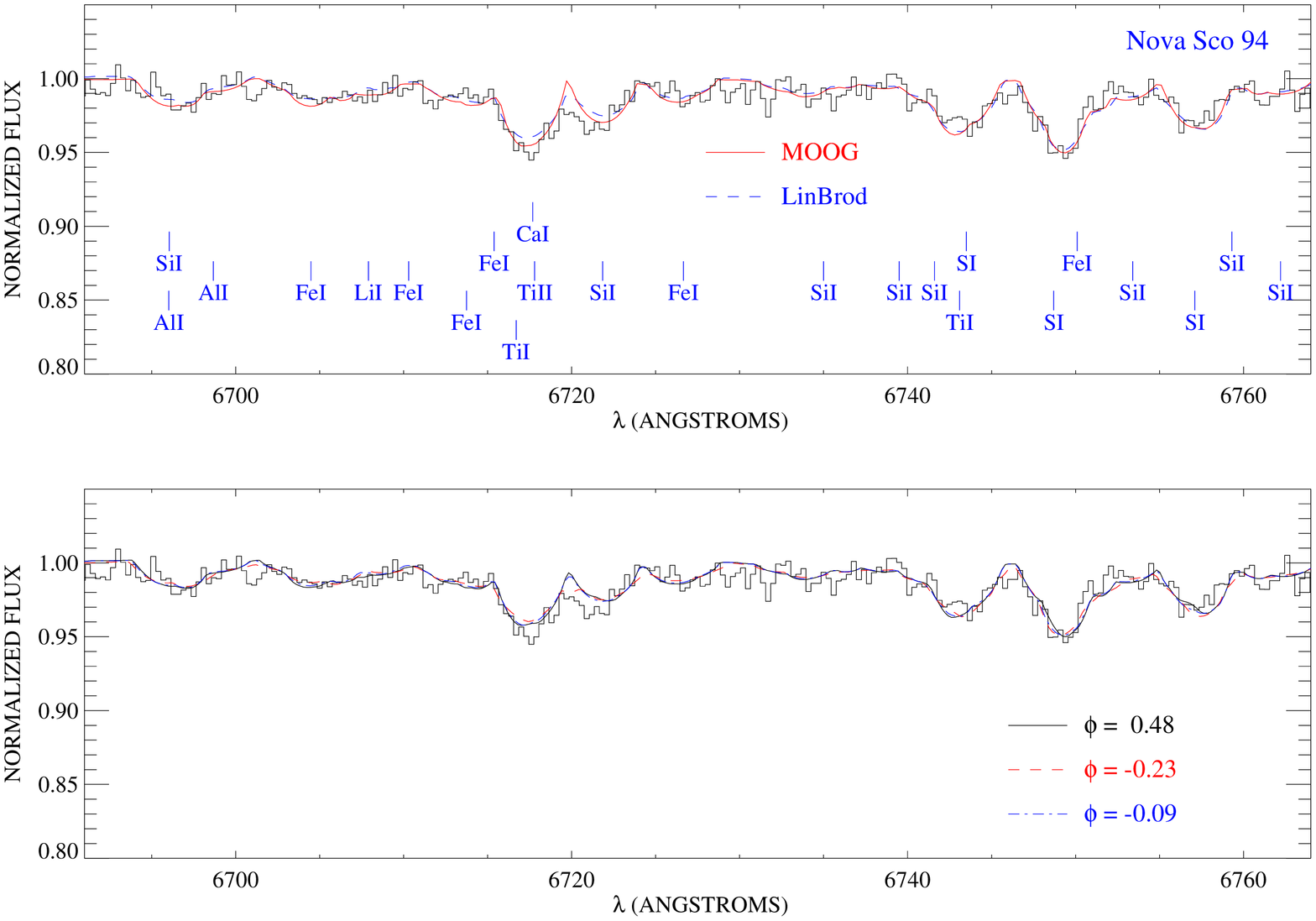}
\caption{\footnotesize{Upper panel: Observed UVES spectrum of the
secondary star in comparison with the average synthetic spectrum
computed with {\scshape LinBrod} (dashed line) for all orbital phases
and the synthetic spectrum computed with MOOG (solid line). The same
element abundances have been adopted for all synthetic
spectra. Bottom panel: Synthetic spectra computed with {\scshape
LinBrod} for different orbital phases.}}     
\label{fig10}
\end{figure*}

\subsubsection{Lithium}

The best fit to the \ion{Li}{i} $\lambda6708$ {\AA} feature provides
an LTE abundance of $\log \epsilon(\mathrm{Li})_{\rm LTE}=2.16 \pm
0.21$. We estimated the non-LTE abundance correction for this element,
from the theoretical LTE and non-LTE curves of growth in Pavlenko \&
Magazz\`u (1996). We found $\Delta_\mathrm{NLTE}= -0.03$. Due to the
weakness of the absorption we consider this abundance estimate given
in Table~\ref{tbl2} as an upper limit\footnote{The template star has
$\log \epsilon(\mathrm{Li})_\mathrm{LTE} \sim 3.3, which does not
affect our results$}. 

According to the effective temperature of the secondary star
($T_{\mathrm{eff}} = 6100 \pm 200$ K), slightly far from the
\emph{lithium gap} (between $6400-6800$ K, Boesgaard \& Tripicco
1986), this Li abundance is consistent with that of main-sequence F
type disc stars with ages in the range $\sim 1-3 \times 10^9$ yr with
the similar metallicity (e.g. Boesgaard \& Tripicco 1987; Balachandran
1991).

\subsection{Comparison with the results of Foellmi et al.}

In a recent paper, Foellmi, Dall \& Depagne (2007) compared the
results on the element abundances reported by Israelian et al. (1999)
using the UVES spectra reported in this paper, taken in 2004, and
additional VLT/UVES spectra taken in 2006. They confirm an enhanced
oxygen abundance and a relatively low Ti and Ca abundance in the
secondary star which is consistent with our results, but also claim
that other $\alpha$-elements do not show over-abundances in clear
disagreement with our findings for Mg, Si and S. Their conclusions are
based on equivalent widths of different spectral features as well as
the comparison of synthetic spectra to the observed \ion{Mg}{i}b
triplet at 5167--83 {\AA}, the \ion{S}{i} triplet at 9228 {\AA} and
some other features of \ion{Mg}{i}, \ion{Si}{i} and \ion{S}{i} in the
spectral region from 8670 to 8740 {\AA}.  

We have revised the equivalent widths presented in Table 1 of
Foellmi et al. (2007) in our VLT/UVES spectra obtained in 2004 and
found slightly larger values which might be related to the continuum
location. In addition, we find it confusing the system velocity of $\sim
-88$ \kms reported by Foellmi et al. (2006) which they also used in
Foellmi et al. (2007), in comparison with our value of $\sim -167$ \kms 
which is comparable to values previously reported in the literature.
However, they obtained an orbital semiamplitude of $K_2=225\pm10$ \kms
which is less accurate than our value, $K_2=226.1\pm0.8$ \kmso. Their
larger error on the orbital semiamplitude could give rise to large
errors on the radial velocities applied to the individual spectra
which might have slightly smoothed out the stellar lines in the
average spectrum of the secondary star. This might explain why these
authors found a rotational velocity of $v \sin i \sim 94$ \kms
slightly larger than our value of $v \sin i \sim 87$ \kmso.

\begin{figure*}
\centering
\includegraphics[width=12cm,angle=90]{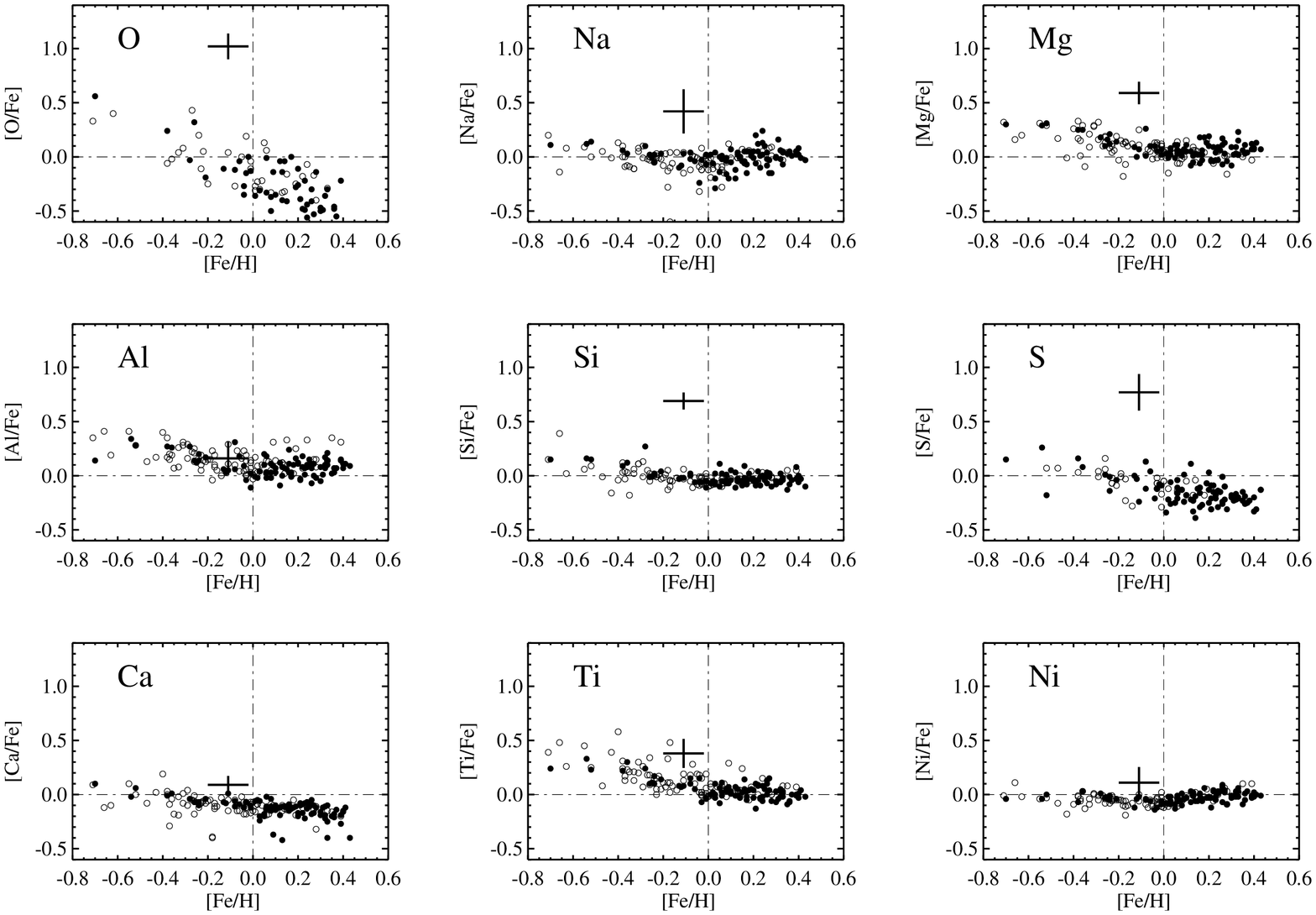}
\caption{\footnotesize{Abundance ratios of the secondary star 
in \mbox{Nova Scorpii 1994} (blue wide cross) in comparison with the
abundances of G and K metal-rich dwarf stars. Galactic trends were
taken from Ecuvillon et al. (2004), Ecuvillon et al. (2006) and Gilli
et al. (2006). The size of the cross indicates the 1-sigma uncertainty. Filled
and empty circles correspond to abundances for planet host stars and
stars without known planet companions, respectively. For the abundance
of oxygen in metal-rich dwarfs we have only considered abundance
measurements in NLTE for the triplet \ion{O}{i} 7771--5 {\AA}. The
dashed-dotted lines indicate solar abundance values.}} 
\label{fig11}
\end{figure*}

\subsubsection{Sulphur}

Foellmi et al. (2007) claim that sulphur is not enhanced in the
secondary star. Indeed, we demonstrate in our paper that the sulphur 
features at 6743--57 {\AA} are very well detected with S/N$\sim150$
(as it is shown in Fig.~\ref{fig8} of this paper) and lead to a large
overabundance of this element. The new features we are referring in
our new paper basically confirm the sulphur overabundance in a
completely independent way to Israelian et al. (1999). Foellmi et al.
did not note these sulphur features in their paper which should be
certainly present in the spectra they used. In addition, in Figs.~1
and~3 in Foellmi et al. (2007), the location of the continuum appears 
improperly low. The \ion{S}{i} features at 8693 and 9228 {\AA} would
lead to a larger abundance if the continuum is suitably placed. We
should remark that due to the large rotational broadening of the
secondary star, roughly 87 \kmso, slight changes of the continuum
location could give rise to considerably differences in the derived
element abundances. The inspection of the lower panel of Fig.~1 in
Foellmi et al. (2007) shows that the \ion{Fe}{i} features, which are
not expected to be enhanced, are not well reproduced by the synthetic
spectrum because the continuum has been placed too low. After
correction of the continuum in this figure, the \ion{S}{i} and
\ion{Si}{i} features at 8680 and 8694 {\AA} would require higher
abundances.

\subsubsection{Silicon}

Foellmi et al. (2007) also argue that silicon is not enhanced.
We should note that we used 12 features of \ion{Si}{i} in our
analysis, and 10 of these features are located in regions of the UVES 
spectrum were the signal-to-noise ratio is roughly 150. We should also
remark that Mg, Si and S show an abundance dispersion from the
measurement of different features of $\sim0.2-0.3$ dex, which indicates
that there are several features which provide lower abundances than
the average which is given in Table~\ref{tbl4}. It is therefore
convenient to derive the element abundances from different features,
and then provide the mean and the dispersion. Foellmi et al. (2007)
have just compared one synthetic spectrum with the observations
without treating each feature individually. It is therefore not
straightforward to work out the reasons for the differences. We should
remark that we have carried out a differential analysis with respect
to the template star and the sun. Thus our atomic data have been
checked by comparing the solar atlas with a synthetic spectrum
computed with the same code and adopting the solar atmospheric
parameters and abundances. Foellmi et al. do not provide information
on their atomic data. Furthermore, we treated carefully the
normalization of the observed spectrum in comparison with the spectrum
of a template star with similar stellar parameters and known
abundances. Thus, we only used as abundance indicators those features
which were well reproduced in the template star. Moreover, it is also 
clear there are some aspects that are difficult to explain in the data
as the strong \ion{Si}{ii} lines in the optical spectrum (as it is
shown in Fig.~\ref{fig5} of this paper), which would require an
overabundance of [Si/H]$\sim 1.75$ dex in LTE. We noted but did not use
these features in our analysis which would add for overabundance. 
We should note that decreasing the surface gravity by 0.2 dex, which
might be expected from the size of the Roche lobe (see
\S~\ref{smgib}), does not improve significantly the agreement
between the synthetic and the observed \ion{Si}{ii} lines. In
addition, the \ion{Fe}{ii} $\lambda6147$ {\AA} feature is not either
well reproduced using this value of the surface gravity, despite that
\ion{Fe}{ii} lines are not expected to show deviations from LTE. 

\subsubsection{Magnesium\label{smgib}}

\begin{figure*}
\centering
\includegraphics[width=9cm,angle=0]{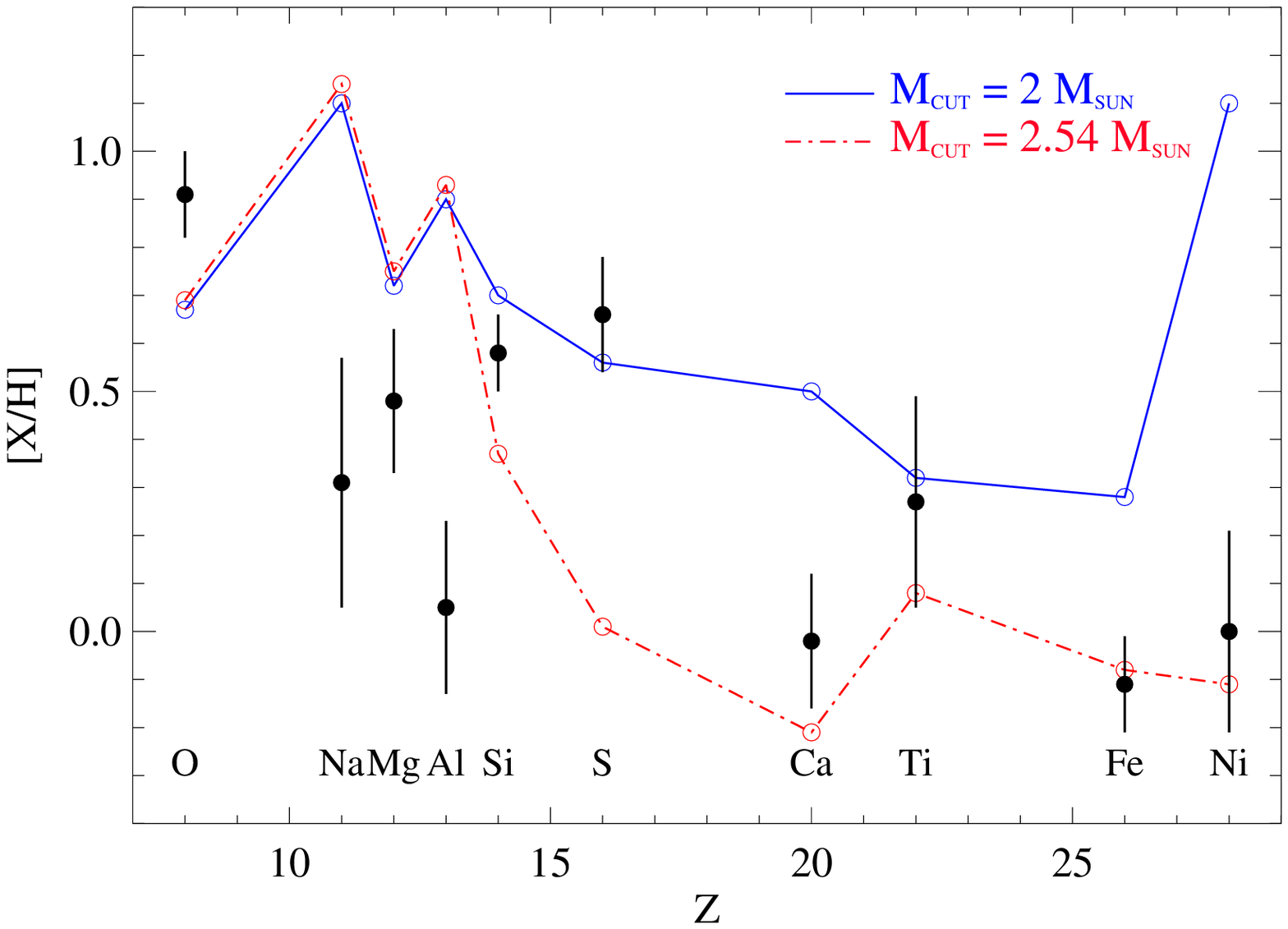}
\includegraphics[width=9cm,angle=0]{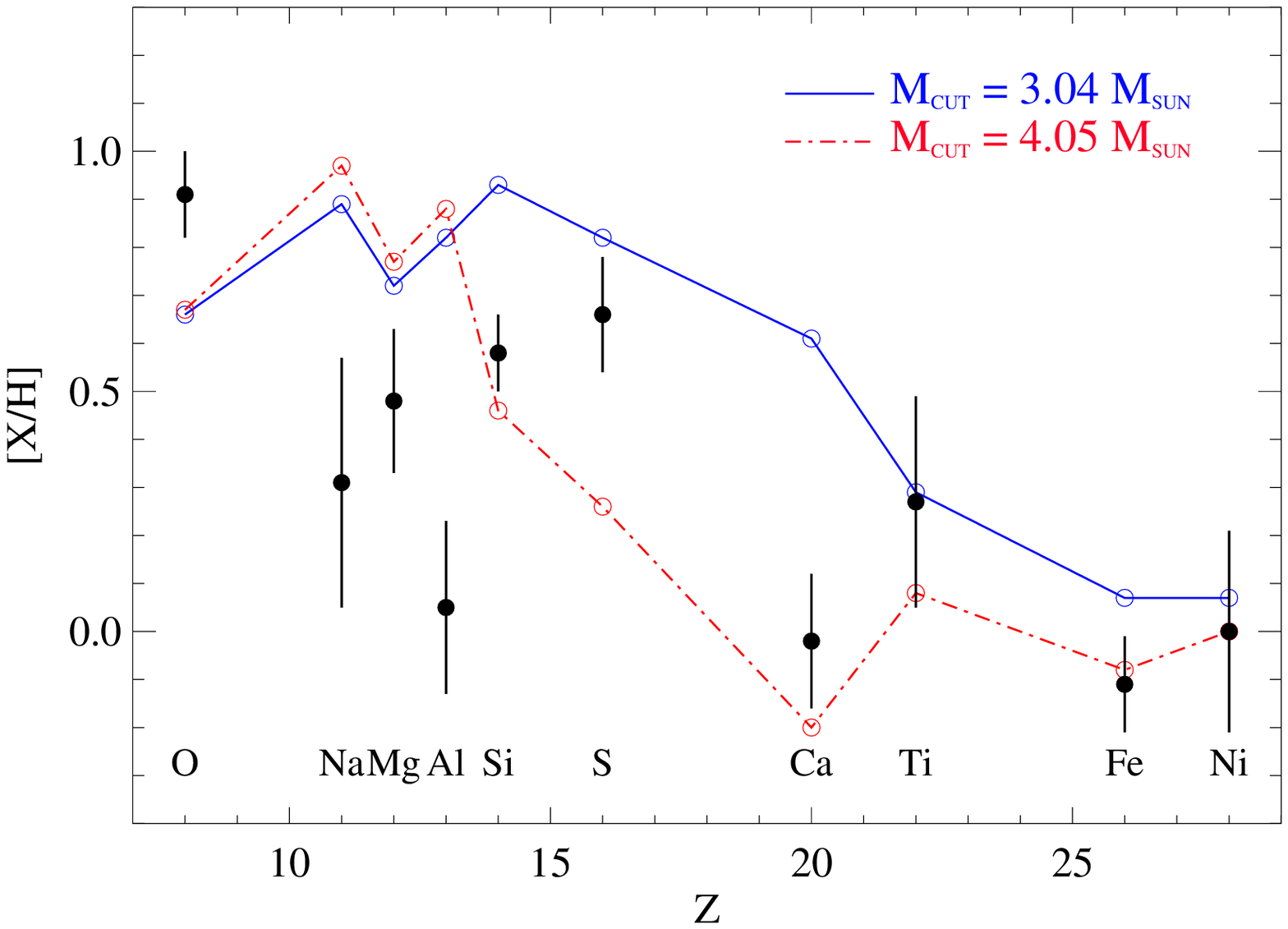}
\caption{\footnotesize{Left panel: Observed abundances (filled circles with the
error bars) in comparison with the expected abundances in the
secondary star after having captured the 70\% of the matter ejected  
within the solid angle subtended by the secondary from a spherically
symmetric supernova explosion of kinetic energy $E_K = 10^{51}$
ergs for two different mass-cuts, $M_{\rm cut}=2$ \Msun (solid
line with open circles), and $M_{\rm cut}=2.54$ \Msun (dashed-dotted
line with open circles). Right panel: the same as left panel but for
a spherically symmetric hypernova explosion of kinetic energy $E_K =
20\times10^{51}$ ergs for two different mass-cuts, $M_{\rm cut}=3.04$
\Msun (solid line with open circles), and $M_{\rm cut}=4.05$ \Msun
(dashed-dotted line with open circles).}}  
\label{fig12}
\end{figure*}

Finally, Foellmi et al. (2007) also discussed the apparently
weakness of the \ion{Mg}{i}b triplet at 5167--83 {\AA}. They infer a
Mg abundance in LTE of [Mg/H]$\sim 0.1$ dex. Our spectroscopic
determination of the abundance provided by the \ion{Mg}{i}b triplet
using the code MOOG is consistent with this result (see
Table~\ref{tbl2}). As discussed in \S~\ref{smg}, the \ion{Mg}{i}b is
quite sensitive to possible errors on the stellar parameters, so we
prefer not to rely on a magnesium determination based just on this
feature. We show in what follows that the Mg Ib triplet is also
sensitive to deviations from sphericity in the stellar atmosphere.
The secondary star is filling its Roche lobe, and therefore is
elongated towards the inner Lagrange point from which the star loses
matter onto the black hole. Thus, the temperature, surface gravity
and Doppler shift at line formation are not constant over the whole
surface of the star, and consequently, the profile of certain lines
depends on the orbital phase. In order to see if this is 
the case for a given spectral feature we have used the program
{\scshape LinBrod} which computes synthetic spectra for each
orbital phase taking into account the Roche lobe symmetry of the
stellar surface (Bitner \& Robinson 2006). The input parameters of
this program are the average effective temperature of the star, which
was assumed to be our spectroscopic estimate of the effective
temperature, and the orbital parameters. We also adopted an orbital
inclination of $\sim69^\circ$ (Beer \& Podsiadlowski 2002) and mass
ratio of $q\sim0.419$ (Shahbaz 2003). The code LinBrod divides the
stellar surface in tiles, for which a local temperature and gravity is
derived, uses the code MOOG to compute a synthetic spectrum for each
tile, and finally integrates over the whole stellar surface. The
average profile of \ion{Mg}{i}b lines over all the observed orbital
phases computed with {\scshape LinBrod} is weaker than the spherical
profile computed with the code MOOG (see Fig.~\ref{fig9}). The main
reason of this difference is that the average surface gravity derived
by {\scshape LinBrod}, which is shared by almost the 80 per cent of
the tiles and only depends on the dynamical parameters, is $\log g =
3.5\pm0.1$, being lower that our estimate ($\log g = 3.7\pm0.2$) but 
marginally consistent within the error bars. The \ion{Mg}{i}b lines
at 5172 {\AA} and 5183 {\AA} become weaker for lower values of $\log
g$. The \ion{Mg}{i}b line at 5167 {\AA} is blended with lines of
other elements that have different sensitivity to changes of the stellar
parameters and they somehow compensate the sensitivity of the
\ion{Mg}{i}b 5167 {\AA} line. This figure also shows how the profile
depends on the orbital phase, mostly due to the different velocity
profile provided by the asymmetry of the stellar surface seen by the
observer.  

In Fig.~\ref{fig9}, we adopted an abundance of [Mg/H]$\sim 0.3$
dex which is almost able to reproduce the observed features when the
code {\scshape LinBrod} is used. Note that we have taken into account
a veiling factor of 7\% estimated at this wavelength according to our
determination of stellar and veiling parameters (see \S~\ref{stam}).
This abundance would be in agreement with the \ion{Mg}{i} feature at
6319 {\AA} which provides [Mg/H]$\sim 0.3$ using the code MOOG and is
less sensitive to the surface gravity (see Table~\ref{tbl2}). 
 
In Fig.~\ref{fig10} we displayed another spectral region where the
\ion{S}{i} 6743--57 {\AA} are located. In this case, in contrast with
the \ion{Mg}{i}b lines, the synthetic spectrum computed with {\scshape
LinBrod} is practically similar to that computed with MOOG, which
indicates that these \ion{S}{i} lines are not affected by the
asymmetry of the stellar surface because they are hardly sensitive to
the surface gravity. The Mg features analysed in the near-IR Keck
spectrum are equally not affected because they are not sensitive to
small variations of the stellar parameters and in addition, this
spectrum was taken at an orbital phase close to 0, when the star
looks almost spherical. 
 
We should remark that in the average Mg abundance given in
Table~\ref{tbl4} we do not applied any correction to the Mg 
abundances derived from the \ion{Mg}{i}b lines using the code MOOG,
which are given in Table~\ref{tbl2}. The use of the program
{\scshape LinBrod} would provide a slightly higher average Mg
abundance and slightly smaller dispersion.  

\section{Discussion}

\subsection{Heavy elements}

The abundances ratios of O, Mg, S, Si and Na with respect to Fe 
in the secondary star of \mbox{Nova Sco 94} are considerably higher
than those in stars of similar Fe content (see Fig.~\ref{fig11}) 
while, on the contrary, Al, Ca, Ti and Ni seem to be consistent with
the Galactic trends. As can be seen in Table~\ref{tbl4}, the
uncertainties induced by effective temperature and gravity are
considerably diminished when dealing with abundance ratios and the
major source of error in $\mathrm{[X/Fe]}$ is associated with the 
dispersion, $\Delta_{\sigma}$, of abundances obtained from different
features of the same element. Thus, the error bars of the abundance
ratios displayed in Fig.~\ref{fig11} were estimated as: 
$$(\Delta[{\rm X/Fe}])^2 = \Delta_{\sigma,{\rm X}}^2 +
(\Delta_{T_{\mathrm{eff}},{\rm X}}-\Delta_{T_{\mathrm{eff}},{\rm
Fe}})^2 +$$ $$(\Delta_{\log g,{\rm X}}-\Delta_{\log g,{\rm
Fe}})^2+\Delta_{\sigma,{\rm Fe}}^2$$
  
Podsiadlowski et al. (2002) showed that a relatively simple model of
pollution of the secondary star from a supernova explosion can explain
most of the observed $\alpha$-element abundance enhancements in the
secondary star.  

Since we have extended the number of elements analysed in the
secondary star, we will intend to extract more information on the SN
explosion by comparing our new results with yields from current
supernova models kindly provided by Umeda, Nomoto et al. (2002, 2005).

\begin{figure*}
\centering
\includegraphics[width=9cm,angle=0]{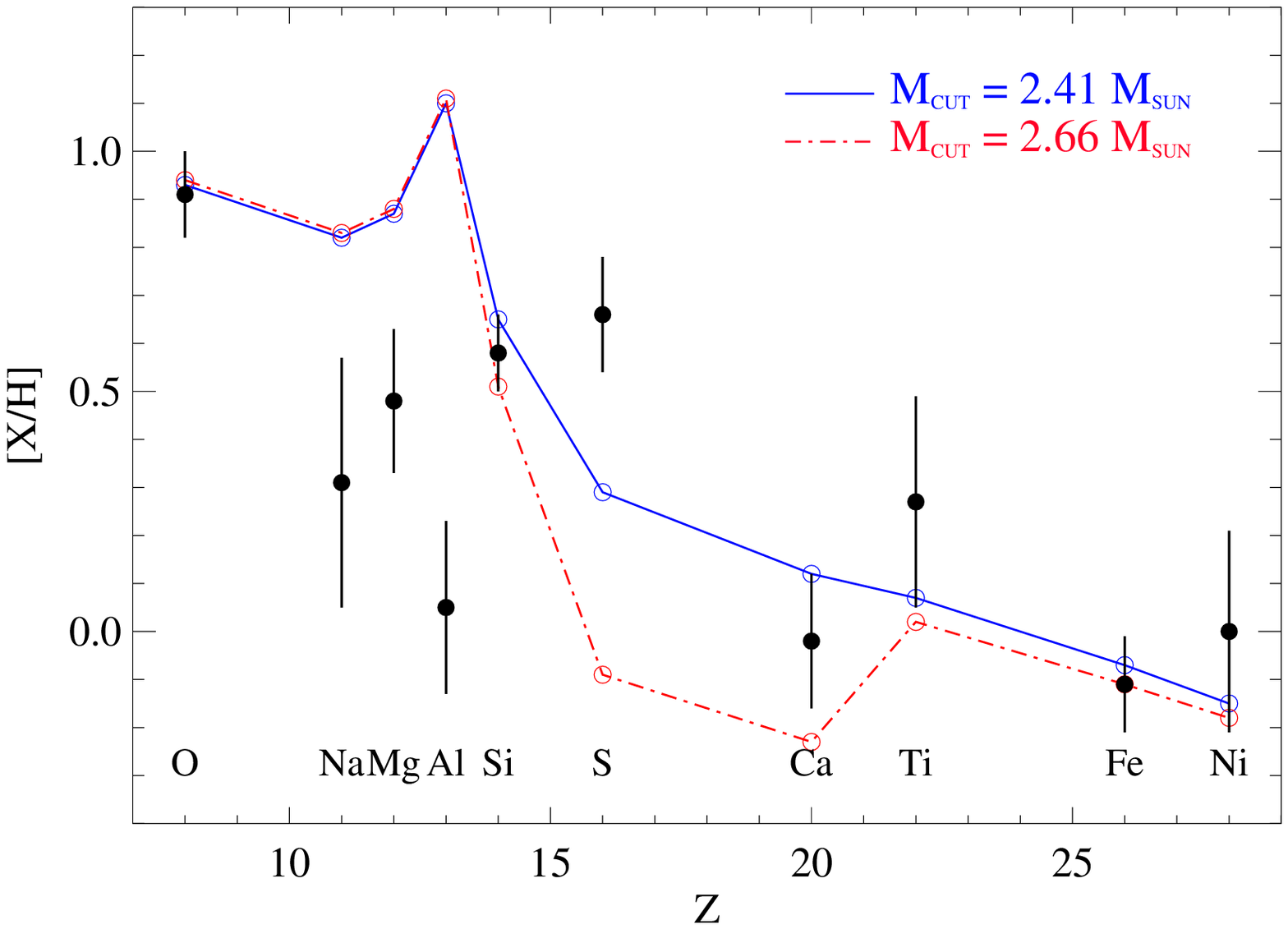}
\includegraphics[width=9cm,angle=0]{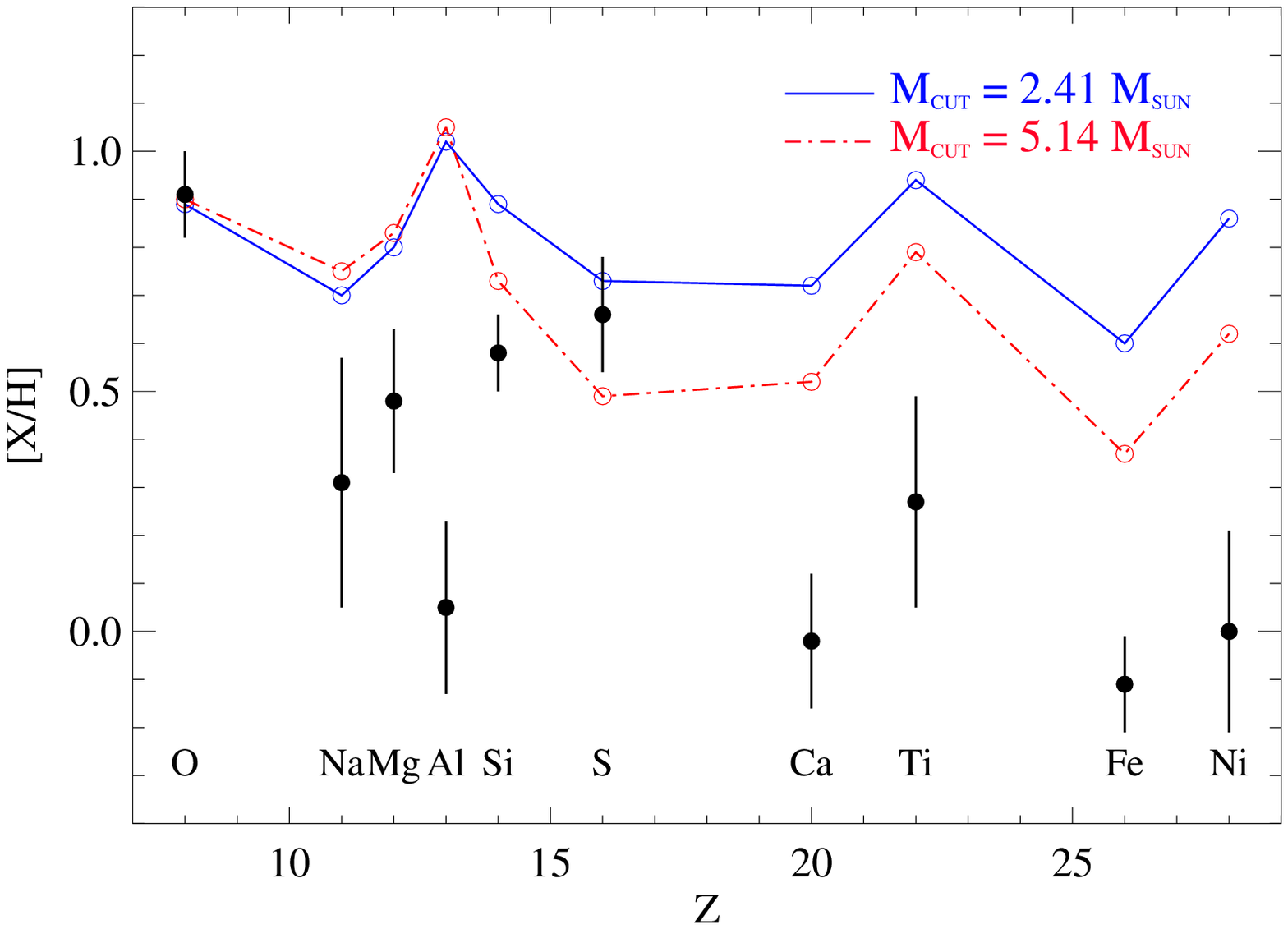}
\caption{\footnotesize{Left panel: Observed abundances (filled circles with the
error bars) in comparison with the expected abundances in the
secondary star after having captured the 10\% of the matter ejected  
within the solid angle subtended by the secondary from a non-spherically
symmetric supernova explosion of kinetic energy $E_K = 10^{52}$
ergs for two different mass-cuts, $M_{\rm cut}=2.41$ \Msun (solid
line with open circles), and $M_{\rm cut}=2.66$ \Msun (dashed-dotted
line with open circles). This model corresponds to the matter ejected
in the equatorial plane of the primary where we assumed that the
secondary star is located (see more details in Gonz\'alez Hern\'andez
et al. 2005b). Right panel: the same as left panel but in this model
we have assumed complete lateral mixing where all the material within
given velocity bins is assumed to be completely mixed. Two simulations
are shown for two different mass-cuts, $M_{\rm cut}=2.41$ \Msun (solid
line with open circles), and $M_{\rm cut}=5.14$ \Msun (dashed-dotted
line with open circles).}}  
\label{fig13}
\end{figure*}

\subsubsection{Spherical SN explosion} 

We have considered spherically symmetric supernova and
hypernova models for a 30 \Msun progenitor (8.5 He core, Umeda \&
Nomoto, 2002, 2005; Tominaga, Umeda \& Nomoto 2007)
and a $\sim 2$ \Msun secondary star, initially placed at an orbital
distance of $\sim 10$ {$R_\odot$} (after the binary orbit has been 
recircularized), which needs to capture almost 70\% of the
matter ejected within the solid angle subtended by the secondary as
seen from the helium star to achieve the observed abundances. We
adopted an initial black hole mass of $M_{\rm BH,i} \sim 5$ \Msun,
before the secondary star begins to transfer matter onto the compact
object. Note that the dynamical estimate of the black hole mass
provides $M_{\rm BH,f}\sim5.4$ \Msun (Beer \& Podsiadlowski 2002). We
have also assumed different mass-cuts ($M_{\rm cut}$, the mass that
initially collapsed forming the compact remnant) and all the fallback
material ($M_{\rm fall}$, amount of mass which is eventually accreted
by the compact core) to be completely mixed with the ejecta. We have 
performed simulations of the expected abundances of the secondary star
after capturing matter from the SN ejecta as in Gonz\'alez Hern\'andez
et al. (2004). In our simulations we varied the mass-cut from 1.35 to
5.05 \Msun in steps of $\sim 0.5$ and 1 \Msun (depending on the mass
bins given in the explosion models), and all the simulations keep
$M_{\rm BH,i}=M_{\rm cut}+M_{\rm fall}$. In addition, all the matter
captured is completely mixed with the whole material of the secondary
star. In Fig.~\ref{fig12} we show the observed abundances in
comparison with the expected abundances for two different mass-cuts.
The error bars in Fig.~\ref{fig12} correspond to the total errors
given in Table~\ref{tbl4}. Note that some of the elements are highly
sensitive to mass-cuts, for instance, the supernova model (left panel
in Fig.~\ref{fig12}) predicts a strong dependence of Ni, Ca and S
abundances on the selected mass-cuts since the explosive burning is
only able to create these elements at relatively deep layers in the
explosion. It would be acceptable to find an agreement between the
observed abundances and one of the two models displayed or at least to
find the observed values in a middle location between both
predictions. This comparison using both spherically symmetric
supernova and hypernova models allow us to extract the following
conclusions:   

\begin{itemize}

\item[a)] The relatively low abundances of Ca, Ti, Ni and Fe may
indicate that the mass-cut was in the range 2--3.5 \Msun for explosion
energies $(1-20)\times10^{51}$ ergs. 

\item[b)] These explosion models suggest that the higher the energy
the larger amounts of elements like S, Si, Ti and Ca, and even Fe and
Ni are created. Thus, the substantial enhancement of Si and S in the
secondary star favours higher explosion energies and hence, the
hypernova scenario (Brown et al.\ 2000). 

\item[c)] In addition, even for higher energies, the amount of S
decreases significantly above a mass-cut of 3.5 \Msun confirming the
need of efficient mixing processes between the fallback matter and the
ejecta (Kifonidis et al.\ 2000).  

\item[d)] The abundances of O, Mg, Al and Na hardly depend on the
explosion energy and mass-cut, but interestingly Al and Na must be
more overabundant than O and Mg in contrast with the observed
abundances. 

\end{itemize}

\subsubsection{Aspherical SN explosion}

The derived heliocentric radial velocity of $\sim -167.1$ \kms is 
much larger than the system kick ($\sim 50$ \kmso) gained from a
spherically symmetric supernova explosion. Brandt et al. (1995)
suggested the black hole in \mbox{Nova Sco 94} could have formed in a
two-stage process where the collapse first leads to the formation of a
neutron star accompanied by a supernova kick, which subsequently
converted into a black hole by fallback. Podsiadloski et al. (2002)
also investigated the expected abundances in the companion star after
pollution of nucleosynthetic products in a non-spherically symmetric
supernova. They found good fits to the observed abundances reported in
Israelian et al. (1999). In their simulations they also took into
account possible losses of mass of the He core progenitor in a
stellar wind before the explosion and the pollution because of the
capture of wind material. This allowed them to use He core progenitor
with masses as high as 16 \Msun, hence reducing the matter ejected in
the SN explosion and preventing from the disruption of the system. 

We have also inspected an aspherical explosion model of a 16 \Msun He
core progenitor with an explosion energy of $10\times10^{51}$ ergs
from Maeda et al. (2002). We did not consider mass losses from the
helium star via stellar winds and hence the greater amount of ejected
matter only requires to capture roughly 10\% of the matter ejected
within the solid angle subtended by the secondary. We computed the
expected abundances in the secondary star after pollution from the SN
products assuming that the secondary star was located in the
equatorial plane of the helium star before the explosion. In this
case, the mass-cut varies from 2.41 to 5.14 \Msun in steps of $\sim
0.6$ \Msun. Fig.~\ref{fig13} shows the comparison of the observed
abundances with the expected abundances in the secondary star after
pollution from the SN ejecta. The error bars in Fig.~\ref{fig13} correspond to the total errors
given in Table~\ref{tbl4}. As in the spherical case we show two
models and two different mass-cuts. The left panel reflects the
composition of the material ejected in the equatorial plane while in
the right panel we have considered complete lateral mixing
(Podsialowski et al.\ 2002), i.e. the ejected matter is completely
mixed within each velocity bin. Note that in the left panel we
selected two close values of the mass-cut while in the right panel the
whole dependence on the mass-cut is shown. The model with complete
lateral mixing tends to enhance all the element abundances for each
mass-cut. The analysis of the aspherical case provides the following
conclusions:    

\begin{itemize}

\item[a)] The aspherical model predicts too low abundances of S
for any mass-cut unless complete lateral mixing is assumed.

\item[b)] However, in the model with complete lateral mixing, in spite
of the fact that Na abundance decreases until being almost marginally
consistent with the observed value, the observed Al still remains too
low in comparison with its expected abundance. 

\item[c)] Moreover, this model increases Ca, Ti, Fe and Ni abundances
by a factor of 2--3 above the observed values. 

\end{itemize}

A complete grid of models with a wide range of progenitor masses and
explosion energies may be required to study in detail the abundance
pattern of the secondary star in this system.

\section{Conclusions}

We have presented a VLT/UVES high resolution spectroscopy of the black
hole binary \mbox{Nova Sco 94}. The individual spectra of the system
allowed us to derive an orbital period of $P=2.62120\pm0.00014$ days
and a radial velocity semiamplitude of the secondary star of
$K_2=226.1\pm0.8$ \kmso. The implied updated mass function is
$f(M)=3.16\pm0.03$ \Msun, consistent with previous values reported in
the literature. The inspection of the high-quality averaged spectrum
of the secondary star provides a rotational velocity
of $v~\sin~i=87^{+8}_{-4}$ \kmso, and hence a a binary mass ratio
$q=0.329\pm0.047$. The derived radial velocity, $\gamma=-167.1\pm0.6$
\kmso, of the center of mass of the system disagrees, at the
3$\sigma$ level, with previous studies.

We have performed a detailed chemical analysis of the secondary star.
We applied a technique that provides a determination of the stellar
parameters, taking into account any possible veiling from the
accretion disc. We find $T_{\mathrm{eff}} = 6100 \pm 200$ K, $\log
(g/{\rm cm~s}^2) = 3.7 \pm 0.2$, $\mathrm{[Fe/H]} = -0.1 \pm 0.1$, 
and a disc veiling (defined as $F_{\rm  disc}/F_{\rm total}$) of less
than 10\% at 5000 {\AA} and decreasing toward longer wavelengths. 

We have revised the chemical abundances of O, Mg, S, Si, Ti and Fe
already reported in Israelian et al. (1999) and determined new element
abundances of Na, Al, Ca, Ni and Li. The element abundances are
typically higher than solar except for Fe, Ca and Ni, and in some
cases significantly enhanced (e.g. O, Mg, S and Si). The abundance
ratio of each element with respect to Fe were compared with those in
stars with similar iron content of the solar neighbourhood. We confirm
that O, Mg, Si, S and Na are considerably overabundant whereas Al, Ca,
Ti and Ni appear to be practically consistent with the Galactic
trends of these elements. We also report an upper limit of the Li
abundance.

These chemical abundances strongly suggest that the secondary star
captured part of the ejecta from a supernova explosion that originated
the black hole in \mbox{Nova Sco 94}. We have compared these element
abundances with element yields from a variety of SN explosion models
for different energies and geometries. An spherically symmetric
explosion model of 30 \Msun progenitor (with a He core of 8.5 \Msuno)
suggests a mass-cut between 2--3.5 \Msun for kinetic
energies $(1-20)\times10^{51}$ ergs, based on the relatively low
abundances of Ca, Ti, Ni and Fe. The greatly enhanced abundances of Si
and specially S in the secondary star favours higher kinetic
energies and therefore a hypernova explosion, and requires efficient
mixing processes between the fallback matter and the ejecta.

The kinematic properties of the system suggest that a natal kick was
imparted to the compact object at birth due to an asymmetry in the
neutrino emission if a neutron star formed first, and/or an asymmetric
mass ejection. We have also inspected a non-spherically symmetric SN
explosion model with a 16 \Msun He core progenitor, but this model
provides unacceptable fits to the observed abundances because they
require complete lateral mixing and produced too high abundances of
Ca, Ti, Fe and Ni. We have also found relatively low Na and Al abundances
which cannot be explained with the current spherical and aspherical SN
models. 

\begin{acknowledgements}

We are grateful to Hideyuki Umeda, Ken'ichi Nomoto, and Nozomu
Tominaga for kindly sending us their spherically symmetric explosion
models and several programs for our model computations. We also thank
Keiichi Maeda for providing us with his aspherical explosion models,
and for helpful discussions. We would like to thank Martin A. Bitner 
and Edward L. Robinson for kindly sending us the code {\scshape
LinBrod}. We are grateful to Tom Marsh for the use of the {\scshape
MOLLY} analysis package. This work has made use of the VALD database
and IRAF facilities. J. I. ackonowledges support from the EU contract
MEXT-CT-2004-014265 (CIFIST). This work has been also funded by the
Spanish Ministry project AYA2005--05149.  

\end{acknowledgements}


\begin{thebibliography}{}

{\scriptsize

\bibitem[Abia \& Mashonkina 2004]{aam04}
Abia, C., \& Mashonkina, L. 2004, \mnras, 350, 1127

\bibitem[Allende Prieto et al. 2004]{all04}
Allende Prieto, C., Barklem, P. S., Lambert, D. L., \& Cunha,
K. 2004, \aap, 53, 181

\bibitem[Al-Naimiy 1978]{aln78}
Al-Naimiy, H. M. 1978, \apss, 420, 183

\bibitem[Andrievsky et al. 1995]{and95}
Andrievsky, S. M., Chernyshova, I. V., Usenko, I. A., Kovtyukh, V. V.,
Panchuk, V. E., \& Galazutdinov, G. A. 1995, \pasp, 107, 219

\bibitem[Asplund 2005]{asp05}
Asplund, M. 2005, \araa, 43, 481

\bibitem[Bailyn et al. 1995]{bai95}
Bailyn, C. D., Orosz, J. A., McClintock, J. E., \& Remillard, R. A.
1995, \nat, 378, 157

\bibitem[Balachandran 1991]{bal91}
Balachandran, S. 1991, \memsai, 62, 33

\bibitem[Baum\"uller, D., \& Gehren 1996]{bag96}
Baum\"uller, D., \& Gehren, T. 1996, \aap, 307, 961

\bibitem[Baum\"uller, D., \& Gehren 1997]{bag97}
Baum\"uller, D., \& Gehren, T. 1997, \aap, 325, 1088

\bibitem[Baum\"uller, D. et al. 1998]{bag98}
Baum\"uller, D., Butler, K., \& Gehren, T. 1998, \aap, 338, 637

\bibitem[Beer \& Podsiadlowski 2002]{bap02}
Beer, M. E., \& Podsiadlowski, Ph. 2002, \mnras, 331, 351

\bibitem[Bitner \& Robinson 1986]{bar06}
Bitner, M. A., \& Robinson, E. L. 2006, \aj, 131, 1712

\bibitem[Boesgaard \& Tripicco 1986]{bat86}
Boesgaard, A. M., \& Tripicco, M. J. 1986, \apj, 302, L49

\bibitem[Boesgaard \& Tripicco 1987]{bat87}
Boesgaard, A. M., \& Tripicco, M. J. 1987, \apj, 313, 389

\bibitem[Brandt et al. 1995]{bra95}
Brandt, W. N., Podsiadlowski, Ph., \& Sigurdsson, S. 1995, \mnras,
277, L35

\bibitem[Brown et al. 2000]{bro00}
Brown, G. E., Lee, C.-H., Wijers, R. A. M. J., Lee, H. K., Israelian, G., \&
Bethe, H. A. 2000, NewA, 5, 191

\bibitem[Buxton \& Vennes 2001]{bav01}
Buxton, M., \& Vennes, S. 2001, Publ.~Astron.~Soc.~Aust., 18, 91

\bibitem[Castelli \& Kurucz 2003 ]{cak03} Castelli, F., \& 
Kurucz, R.~L.\ 2003, IAU Symposium, 210, 20P 

\bibitem[Ecuvillon et al. 2004]{ecu04}
Ecuvillon, A., Israelian, G., Santos, N. C., Mayor, M., Villar, V., \&
Bihain, G. 2004, \aap, 426, 619 

\bibitem[Ecuvillon et al. 2006]{ecu06}
Ecuvillon, A., Israelian, G., Santos, N. C., Shchukina, N. G., Mayor,
M., \& Rebolo, R. 2006, \aap, 445, 633

\bibitem[Foellmi et al. 2006]{foe06}
Foellmi, C., Depagne, E., Dall, T. H., \& Mirabel, I. F. 2006, \aap,
457, 249

\bibitem[Foellmi et al. 2007]{foe07}
Foellmi, C., Dall, T. H., \& Depagne, E. 2007, \aap, 464, L61

\bibitem[Galazutdinov et al. 2000]{gal00}
Galazutdinov, G. A., Musaev, F. A., Krelowski, J., \& Walker, G. A. H.
2000, \pasp, 112, 648 

\bibitem[Gilli et al. 2006]{gua06}
Gilli, G., Israelian, G., Ecuvillon, A., Santos, N. C., \&
Mayor, M. 2006, \aap, 449, 723.

\bibitem[Gonz\'alez Hern\'andez et al. 2004]{gon04}
Gonz\'alez Hern\'andez, J. I., Rebolo, R., Israelian, G., Casares, J.,
Maeder, A., \& Meynet, G. 2004, \apj, 609, 988

\bibitem[Gonz\'alez Hern\'andez et al. 2005]{gon05}
Gonz\'alez Hern\'andez, J. I., Rebolo, R., Israelian, G., Casares, J.,
Maeda, K., Bonifacio, P., \& Molaro, P. 2005, \apj, 630, 495

\bibitem[Gonz\'alez Hern\'andez et al. 2006]{gon06}
González Hernández, J. I., Rebolo, R., Israelian, G., Harlaftis, E.
T., Filippenko, A. V., \& Chornock, R. 2006, \apj, 644, L49

\bibitem[Grevesse et al. 1996]{gre96}
Grevesse, N., Noels, A., \& Sauval, A. J. 1996, \emph{The sixth
annual October Astrophysics Conference}, ASP Conf. Ser. 99, 117 

\bibitem[van der Hooft 1998]{hoo98}
van der Hooft, F., Heemskerk, M. H. M., Alberts, F., \& van Paradijs,
J. 1998, \aap, 329, 538

\bibitem[Israelian et al. 1999]{isr99}
Israelian, G., Rebolo, R., Basri, G., Casares, J., \&
Mart{\'\i}n, E. L. 1999, \nat, 401, 142

\bibitem[Kifonidis et al. 2000]{kif00}
Kifonidis, K., Plewa, T., Janka, H.-Th., \& Müller, E. 2000, \aap, 531, L123 

\bibitem[Kurucz et al. 1993]{kur93}
Kurucz, R. L. ATLAS9 Stellar Atmospheres Programs and 2 \kms
Grid. (CD-ROM, Smithsonian Astrophysical Observatory, Cambridge,
1993).

\bibitem[Kurucz et al. 1984]{kur84}
Kurucz, R. L., Furenild, I., Brault, J., \& Testerman, L. 1984, Solar Flux Atlas from
296 to 1300 nm, NOAO Atlas 1 (Cambridge: Harvard Univ. Press)

\bibitem[Maeda et al. 2002]{mae02}
Maeda, K., Nakamura, T., Nomoto, K., Mazzali, P. A., Patat, F., \&
Hachisu, I. 2002, \apj, 565, 405

\bibitem[Marsh et al. 1994]{mah94}
Marsh, T. R., Robinson, E. L., \& Wood, J. H. 1994, \mnras, 266, 137

\bibitem[Mashonkina et al. 2007]{mas07}
Mashonkina, L., Korn, A. J., \& Przybilla, N. 2007, \aap, 461, 261

\bibitem[Mirabel et al. 2002]{mir02}
Mirabel,  I. F., Mignani, R., Rodrigues, I., Combi, J. A., Rodr{\'\i }guez, L. F., \&
Guglielmetti, F. 2002, \aap, 395, 595

\bibitem[Orosz \& Bailyn, 1997]{oab97}
Orosz, J. A., \& Bailyn, C. D. 1997, \apj, 477, 876

\bibitem[Pavlenko \& Magazzu 1996]{pam96}
Pavlenko, Ya. V., \& Magazz\`u, A. 1996, \aap, 311, 961

\bibitem[Piskunov et al. 1995]{pis95}
Piskunov, N. E., Kupka, F., Ryabchikova, T. A., Weiss, W. W., \& Jeffery, C.
S. 1995, \aaps, 112, 525

\bibitem[Podsiadlowski et al. 2002]{pod02}
Podsiadlowski, P., Nomoto, K., Maeda, K., Nakamura, T., Mazzali, P., \&
Schmidt, B. 2002, \apj, 567, 491 

\bibitem[Sbordone et al. 2004]{sbo04}
Sbordone, L., Bonifacio, P., Castelli, F., \& Kurucz, R. L., 2004, 
MSAIS, 5, 93 

\bibitem[Shimanskaya et al. 2000]{shi00}
Shimanskaya, N. N., Mashonkina, L. I., \& Sakhibullin, N. A. 2000,
Astron. Rep., 44, 530

\bibitem[Shchukina 1987]{shc87}
Shchukina, N.G. 1987, Kinemat. Phys. Celest. Bodies, 3, 36

\bibitem[Shchukina \& Trujillo Bueno 2001]{sat01}
Shchukina, N.G., \& Trujillo Bueno, J. 2001, \apj, 550, 970

\bibitem[Shahbaz et al. 1999]{sha99}
Shahbaz, T., van der Hooft, F., Casares, J., Charles, P. A., \& van
Paradijs, J. 1999, \mnras, 306, 89 

\bibitem[Shahbaz 2003]{sha03}
Shahbaz, T. 2003, \mnras, 339, 1031 

\bibitem[Sneden 1973]{sne73}
Sneden, C. 1973,  PhD Dissertation (Univ. of Texas at Austin)

\bibitem[Takada-Hidai 2002]{tak02}
Takada-Hidai, M., Takeda, Y., Sato, S. et al. 2002, \apj, 573, 614

\bibitem[Takeda et al 2005]{tak05}
Takeda, Y., Hashimoto, O., Taguchi, H. et al. 2005, \pasj, 57, 751

\bibitem[Takeda \& Honda 2005]{tah05}
Takeda, Y., \& Honda, S. 2005, \pasj, 57, 65

\bibitem[Tominaga et al. (2007)]{tom07} 
Tominaga, N., Umeda, H. \& Nomoto, K. 2007, \apj, 660, 516

\bibitem[Umeda \& Nomoto 2002]{uan02}
Umeda, H.,\& Nomoto, K. 2002, \apj, 565, 385

\bibitem[Umeda \& Nomoto 2005]{uan05}
Umeda, H.,\& Nomoto, K. 2005, \apj, 619, 427

\bibitem[Wedemeyer 2001]{wed01}
Wedemeyer, S. 2001, \aap, 373, 998

\bibitem[Zhang et al. 1994]{zha94}
Zhang, S. N.,Wilson, C. A., Harmon, B. A., Fishman G. J., Wilson, R.
B. et al.\ 1994, \emph{IAU Circ.}, 6046

\bibitem[Zhao et al. 1998]{zha98}
Zhao, G., Butler, K., \& Gehren, T. 1998, \aap, 333, 219

\bibitem[Zhao et al. 2000]{zha00}
Zhao, G., \& Gehren, T. 2000, \aap, 362, 1077

}

\end{thebibliography}
\end{document}